\documentclass[aps,psfig,prd,preprintnumbers,amsmath,amssymb,superscriptaddress]{revtex4}
\usepackage{graphics}

\begin{document}
\title{Inflationary Dynamics with Runaway Potentials}
\author{P.R.~Ashcroft}
\affiliation{Department of Applied Mathematics and Theoretical
Physics, Centre for Mathematical Sciences, University of Cambridge,
Wilberforce Road , Cambridge CB3 OWA, UK}
\author{C. van de Bruck}
\affiliation{Astrophysics Department, Oxford University, Keble
Road, Oxford OX1 3RH, UK}
\author{A.-C. Davis}
\affiliation{Department of Applied Mathematics and Theoretical
Physics, Centre for Mathematical Sciences, University of Cambridge,
Wilberforce Road , Cambridge CB3 OWA, UK}
\date{19 September 2003}
\begin{abstract}
The evolution of two slow--rolling scalar fields with 
potentials of the form $V=V_0 \phi^{-\alpha}\exp(-\beta \phi^m)$ 
is studied. Considering different values of the parameters 
$\alpha$, $\beta$ and $m$, we derive several new inflationary 
solutions in which one fields just evolves in the background 
and is not important for the inflationary dynamics. 
In addition, we find new solutions where both fields 
are dynamically important during inflation. 
Moreover, we discuss the evolution of perturbations in both scalar fields and
the spacetime metric, concentrating on the production of entropy 
perturbations. We find that for a large region 
in parameter space and initial conditions tensor modes are negligible
and that adiabatic and isocurvature perturbations are essentially 
uncorrelated.  
\end{abstract}

\maketitle 
\vspace{0.5cm}
\noindent DAMTP-2002-125

\section{Introduction}
Since their advent just over twenty years ago, cosmological inflationary
models have been developed to try to deal with several problems
arising in the standard cosmology. The most important consequence of 
an inflationary stage in the very early universe is the development of 
perturbations in space--time and matter, which eventually evolve into
the structures we observe in the universe today. The simplest
model with an inflationary stage is where inflation is 
driven by a single scalar field, called the inflaton, with some 
potential $V(\phi)$. If the potential is flat enough, theory predicts
that adiabatic perturbations are generated which obey the Gaussian 
statistics and have an almost scale-free spectrum (for recent reviews see
e.g. \cite{liddlelyth} and \cite{Riotto}). 

However, the most important drawback for inflationary cosmology is that 
there is no unique candidate for the inflaton field and its potential. 
Furthermore, according to our theories of particle physics, there 
are a large number of scalar fields which, potentially, could be important in 
the early universe. For example, hybrid inflation is a model, in which 
two scalar fields are important: whereas one field drives an
inflationary epoch, the dynamics of the second field will end this 
period of inflation \cite{hybrid}. If inflation itself is driven 
by two or more scalar fields, the perturbations are no longer 
purely adiabatic or Gaussian \cite{Kofman}-\cite{bartolo}.  
Moreover, the transition to the normal radiation--dominated epoch 
depends on the decay of the scalar fields into radiation and matter,
which could influence the evolution of perturbations as well. 
Cosmological observations by the Wilkinson Microwave Anisotropy Probe (WMAP), 
the Planck Surveyor, 2dF and SLOAN Digital Sky Survey will put strong 
constraints on the nature of the primordial perturbations and
therefore put constraints on more complicated models of inflation 
and the process of reheating. 

Although formalisms have been presented in order to study two--field 
inflation (see e.g. \cite{gordon}, \cite{bartolo} and \cite{wands}), no explicit model
has been used in order to examine the background evolution of both
fields and the consequences for the perturbations. In this paper, 
we investigate the evolution of two scalar fields, 
each with a potential of the form $V(\phi) = V_0 \phi^{-\alpha} 
\exp[ -\beta \phi^m]$, and examine the effect of the parameters 
$\alpha, \beta, m$ on the evolution of the universe. This potential 
includes pure exponentials ($\alpha = 0$), pure power 
law ($\beta = 0$) and combinations of the two. For a single scalar
field, this potential was investigated in detail in \cite{barrow} and 
\cite{parsons}. Exponential potentials 
appear in Kaluza--Klein theories as well as in supergravity and 
superstring models (for a review, see e.g. \cite{Pedro}). 
Models based on dynamical supersymmetry 
breaking involving fermion condensates motivate inverse power 
potentials \cite{binetruy}; supergravity corrections to these models in turn 
predict potentials of the form above \cite{brax}. Furthermore, models 
motivated by higher--dimensional theories and/or scalar--tensor
theories easily incorporate these potentials. For example, the form of
the potential was used in \cite{damour} to examine inflation with a
background dilaton field.

The potentials just described 
were used both in inflationary cosmology as well as in models 
for dark energy, driving the observed acceleration of the universe 
at the present epoch (for a review, see \cite{ratra}). 
Indeed, the potential above has interesting 
properties, allowing for scaling behavior and other attractor--like 
solutions (see e.g. \cite{brax2} and \cite{ng}). 
However, our goal in this paper is to obtain an
understanding of the dynamics of two scalar fields with this potential which drive an inflationary epoch in the {\it early} 
universe. This, in turn, allows us to study further the consequences for 
the perturbations, in the two scalar fields and the space--time metric, 
{\it during} the inflationary epoch. 
This can be seen as a first step to understanding the initial
perturbations in the radiation dominated era, which eventually 
evolve into the structures we observe today. We would 
like to emphasize, however, that in order to calculate the
perturbations in the radiation dominated epoch, an understanding of
the decay of both scalar fields is needed. There are some possibilities
we would like to mention: 
\begin{itemize}
\item Both scalar fields decay, one into radiation and baryonic
matter, the other field only into dark matter. The second field 
could, in principle, also decay only partly into dark matter, whereas 
the ``rest'' provides an explanation for dark energy. 
\item Only one scalar field decays completely into matter and
radiation, the other decays partly into some form of matter; the 
remains of the field play the role of dark energy. 
\item One of the fields plays the role of the curvaton field 
\cite{curvaton}, i.e. 
it decays well after the inflationary epoch and is responsible for 
the curvature perturbation. 
\end{itemize}
There are more possibilities, of course, but it is clear that the 
subsequent evolution of perturbations in the metric and matter 
fields strongly depends on which of these possibilities is realized. 
The situation is now far more complicated when compared to inflationary 
models based on one scalar field alone. 

The paper is organized as follows: in Section 2 we present the 
equations of motion; section 3 discusses the case of two 
independent potentials for both fields, i.e. the total potential 
is given by $V = V_{0_\phi} \phi^{-\alpha} \exp[ -\beta \phi^m] + 
V_{0_\chi} \chi^{-\alpha} \exp[ -\beta \chi^m]$; in Section 4 we investigate 
a mixed potential of the form $V = V_0 \chi^{-\alpha} \exp[ -\beta
\phi^m]$. In sections 3 and 4 we discuss all of the possible cases  in detail. These
two sections are necessarily technical. However, we have included a
summary of our results at the end of each section. Readers who are not
interested in the technical aspects may wish to refer to these
summaries. The perturbations are discussed in Section 5; we present our 
conclusions in Section 6.

\section{The Field Dynamics}
Our starting point shall be that of Einstein gravity with two scalar
fields and a general potential $V(\phi,\chi)$. The action for
this (in Planck units) is described by
\begin{eqnarray}
\mathcal{S} = \int \mathrm{d^4x} \sqrt{-g} \left[\frac{1}{2} R -
\frac{1}{2}  (\partial_a \phi \partial^a \phi + \partial_a \chi
\partial^a \chi ) - V(\phi , \chi) \right].
\end{eqnarray}
This action leads to the usual Friedmann equation and two
Klein--Gordon equations for both scalar fields:
\begin{eqnarray}
H^2 &=& \frac{1}{3} \left[ \frac{1}{2} \left(\dot{\phi}^2 + \dot{\chi}^2\right) + V(\phi,\chi)
\right],\nonumber \\
\ddot{\phi} &=& -3H\dot{\phi} - V_\phi,\label{eq:evolution} \\
\ddot{\chi} &=& -3H\dot{\chi} - V_\chi,\nonumber
\end{eqnarray}
where we use the notation $V_\phi = \partial V / \partial \phi$ with
its obvious extensions.

We shall find it convenient to make the slow--roll approximation which
is appropriate when the slow--roll parameters
\begin{eqnarray}
\epsilon_{\phi} = \frac{1}{2} \left( \frac{V_{\phi}}{ V}
\right)^2, \ \ \ \epsilon_{\chi} = \frac{1}{2} \left( \frac{V_{\chi} }{ V}
\right)^2, \nonumber \\
\eta_{\phi \phi} = \frac{V_{\phi \phi}}{V},\ \ \eta_{\phi \chi} = \frac{
V_{\phi\chi}}{V}, \ \ \eta_{\chi\chi} = \frac{V_{\chi\chi}}{V}. \label{eq:slowparam}
\end{eqnarray}
satisfy $|\epsilon_i, \eta_{ij} | \ll 1, i,j = \phi, \chi$. These five
conditions are sufficient for us to generate an inflationary epoch
where $\ddot{a}(t) > 0$. With this approximation the equations of
motions~(\ref{eq:evolution}) reduce to
\begin{eqnarray}
H^2 &=& \frac{1}{3} V(\phi,\chi),\label{eq:slow1}\\
3H\dot{\phi} &=& - V_\phi,\label{eq:slow2}\\ 
3H\dot{\chi} &=& - V_\chi\label{eq:slow3}.
\end{eqnarray}
For the most part, we shall see that slow--roll is an attractor in our
model and that we shall be justified in making this approximation.

\section{Uncoupled Potentials}
In this section we begin our investigation by considering two 
uncoupled potentials of the form
\begin{eqnarray} 
 V(\phi, \chi) = V_{0_\phi}
 \phi^{-\alpha_1}\exp\left[-\beta_1\phi^{m_1}\right] 
+ V_{0_\chi} \chi^{-\alpha_2}\exp\left[-\beta_2\chi^{m_2}\right]. 
\end{eqnarray}
In this case the slow--roll parameters read 
\begin{eqnarray}
\epsilon_\phi &=& \frac{1}{2}\left(\frac{\alpha_1}{\phi} 
+ \beta_1 m_1 \phi^{m_1 -1} \right)^2 \frac{ V_{0_\phi}^2
 \phi^{-2\alpha_1}\exp\left[-2\beta_1\phi^{m_1}\right] }{(V(\phi,\chi))^2} , \nonumber \\
\eta_{\phi\phi} &=& \left[\left(\frac{\alpha_1}{\phi} + \beta_1 m_1 \phi^{m_1
-1} \right)^2 + \frac{\alpha_1}{\phi^2} + \beta_1^2 m_1 ( m_1 - 1)
\phi^{m_1 - 2} \right]\frac{ V_{0_\phi}
 \phi^{-\alpha_1}\exp\left[-\beta_1\phi^{m_1}\right]}{V(\phi,\chi)} ,
\end{eqnarray}
and similarly for $\chi$, with $\eta_{\phi \chi} = 0$.
It is clear that, with a judicious choice of initial conditions, 
these can always be made small. It is also clear 
that $m_i = 1 $ will be a transitional value because, for $m_i > 1$, 
the slow--roll parameters will grow. We now begin to examine the
evolution of the fields in this regime. We shall assume that $\beta_1
> 0$ but make no further assumptions about the other parameters in the
potential terms. 

\vspace{0.5cm}
\begin{center}
\textbf{Large Field Values}
\end{center}

\subsection{ $m_1 , m_2 > 0$ }
With $ m_1 , m_2 > 0$, the exponential part of the potential, for
large enough $\phi, \chi$ at least, 
will always dominate over the power law. Dividing equations
(\ref{eq:slow2}-\ref{eq:slow3}) we generate the equation
\begin{eqnarray}
\frac{\dot{\phi}}{V_\phi} = \frac{\dot{\chi}}{V_\chi} \label{eq:phichi0}
\end{eqnarray}
We are unable to integrate this directly but are able to do so to
leading order in the fields by means of integration by parts. This
then generates for large $\phi, \chi$
\begin{eqnarray}
\beta_1^2 m_1^2 \phi^{2m_1 - 2}  V_{0_\phi}
 \phi^{-\alpha_1}\exp\left[-\beta_1\phi^{m_1}\right] =\beta_2^2 m_2^2
 \chi^{2m_2 - 2} V_{0_\chi}
 \chi^{-\alpha_2}\exp\left[-\beta_2\chi^{m_2}\right].\label{eq:firstapprox}
\end{eqnarray}
One then takes natural logarithms to find a simple expression between
the fields. Since $m_i > 0 $ the terms proportional to the powers of
the fields will eventually dominate. This gives us
\begin{eqnarray}
\beta_1 \phi^{m_1} = \beta_2 \chi^{m_2},\label{eq:phichi1}
\end{eqnarray}
to leading order. Then using (\ref{eq:slow1}) and (\ref{eq:phichi1})
we are able to show
\begin{eqnarray}
H(\phi) = \left(\frac{V_{0_\phi}}{3} \right)^{\frac{1}{2}}
\phi^{-\frac{\alpha_1}{2}}
\exp \left[ \frac{-\beta_1 \phi^{m_1}}{2} \right] \left( 1 +
\frac{m_1^2}{m_2^2} \left( \frac{\beta_1}{\beta_2}
\right)^{\frac{2}{m_2}} \phi^{\frac{2m_1}{m_2} - 2}
\right)^{\frac{1}{2}}.\label{eq:hub1}
\end{eqnarray}
Substituting back into (\ref{eq:slow2}) we must solve the
differential equation for $\phi$
\begin{eqnarray}
 \dot{\phi} &=&\left( \frac{V_{0\phi}}{3}\right)^{\frac{1}{2}} \frac{
\frac{\alpha_1}{\phi} + \beta_1 m_1 \phi^{m_1 -1}}{\left( 1+ \frac{m_1^2}{m_2^2}\left(
\frac{\beta_1}{\beta_2} \right)^{\frac{2}{m_2}} \phi^{\frac{2m_1}{m_2} - 2}
\right)^{\frac{1}{2}}} 
\phi^{- \frac{\alpha_1}{2}} \exp \left[ \frac{-\beta_1 \phi^{m_1}}{2}
\right], \\
&=&\left( \frac{V_{0\phi}}{3}\right)^{\frac{1}{2}} \frac{ \beta_1
m_1 \phi^{-\frac{\alpha_1}{2} + m_1 -1}}{\left(1+ \frac{m_1^2}{m_2^2} \left(
\frac{\beta_1}{\beta_2} \right)^{\frac{2}{m_2}} \phi^{\frac{2m_1}{m_2} - 2}
\right)^{\frac{1}{2}}} \exp \left[ \frac{-\beta_1 \phi^{m_1}}{2}
\right],\label{eq:dominant}
\end{eqnarray}
to leading order in $\phi$. For large $\phi$, we are able to integrate
this expression under two different regimes, depending on whether $m_1 <
m_2$ or $ m_1 = m_2$. This corresponds to one field giving the dominant
contribution to the Hubble parameter and both giving equal
contributions. This is clear from the denominator in
(\ref{eq:dominant}). We shall take both of these cases in turn.
\subsubsection{ $m_1  < m_2$} \label{sec:firstuncoup}
In this case it is easy to see that the $\phi$ field is the dominant
contributor. Solving for $\phi$ we integrate (\ref{eq:dominant}) by
parts to show
\begin{eqnarray}
t(\phi) = \left( \frac{3}{V_{0_\phi}} \right)^{\frac{1}{2}}
\frac{2}{\beta_1^2 m_1^2} \phi^{\frac{\alpha_1}{2} - 2m_1 +2}
\exp\left[ \frac{\beta_1 \phi^m_1}{2}\right]. \label{eq:phi1}
\end{eqnarray}
Immediately we are able to see that large $\phi$ implies large $t$ and
so inflation will occur at late times with this set up. We are unable to invert this directly. However by taking $\ln$, we see
that the exponential quickly dominates, and so to leading order
\begin{eqnarray}
\phi(t) = \left[ \frac{2}{\beta_1} \ln\left[ \left(
\frac{V_{0_\phi}}{3} \right)^{\frac{1}{2}} \frac{\beta_1^2 m_1^2}{2} t
\right] \right]^{\frac{1}{m_1}}. \label{eq:phi2}
\end{eqnarray}
Writing (\ref{eq:phi1}) in the form
\begin{eqnarray}
\frac{\beta_1 \phi^{m_1}}{2} = \ln \left[  \left(
\frac{V_{0_\phi}}{3} \right)^{\frac{1}{2}} \frac{\beta_1^2 m_1^2}{2} \phi^{-\frac{\alpha_1}{2} + 2m_1 -2} t
\right],
\end{eqnarray}
we are able to substitute back into (\ref{eq:phi2}) to obtain a
solution for $\phi(t)$ to next order in $t$. A little thought reveals
why we want and need to do this. When we come to evaluate the Hubble
parameter as a function of time we need to take the exponent of
$\phi$. Every time we do this we must take the next to leading order
expression. This is given by
\begin{eqnarray}
\phi(t) = \left[ \frac{2}{\beta_1} \ln\left[ \frac{ \left(
\frac{V_{0_\phi}}{3} \right)^{\frac{1}{2}} \frac{\beta_1^2 m_1^2}{2}
t } { \left[ \frac{2}{\beta_1} \ln\left[ \left(\frac{V_{0_\phi}}{3} \right)^{\frac{1}{2}} \frac{\beta_1^2 m_1^2}{2} t
\right] \right]^{\frac{4+\alpha_1}{2m_1} - 2 }} \right] \right]^{\frac{1}{m_1}}. \label{eq:phi3}
\end{eqnarray}
Then we plug this into (\ref{eq:hub1}) to solve for $H(t)$ and then
integrate to get the scalefactor which gives us,
\begin{eqnarray}
H(t)& = &\frac{1}{\beta_1 m_1^2} \left( \frac{2}{\beta_1}
\right)^{\frac{2-m_1}{m_1}} \frac{1}{t} \left[ \ln t
\right]^{\frac{2-m_1}{m_1}}, \\
a(t)& \propto& \exp\left[ \frac{1}{\beta_1 m_1 ( 2- m_1)}
\left[ \frac{2}{\beta_1}\ln t
\right]^{\frac{2-m_1}{m_1}}\right].
\end{eqnarray}
In the evolution of the fields we may drop most of the coefficients as
they simply result in constant terms in the solution. This gives the
final solution 
\begin{eqnarray}
\phi(t) &=& \left( \frac{2}{\beta_1} \ln t \right)^{\frac{1}{m_1}},\\
\chi(t) &=& \left( \frac{2}{\beta_2} \ln t \right)^{\frac{1}{m_2}}.
\end{eqnarray}
Now because $m_i < 1$, we see that the scale factor grows faster than
power law. This regime is effectively equivalent to a single scalar
field driving the dynamics of the universe as the second scalar $\chi$
sits in the background. Unsurprisingly, the solutions are in direct
agreement with section III.A of \cite{parsons}.

\subsubsection{  $m_1  > m_2$}
This is equivalent to the previous section by symmetry.

\subsubsection{ $m_1 = m_2 = m$}
We now consider the case where both the fields give equal contributions to the
evolution of our universe.  Since, the second field is now
significant,  we
expect to see something new. 
This case includes that of two straight
exponential potentials which has already been studied in
\cite{coley}. We are able to integrate (\ref{eq:dominant}) to
leading order to show that
\begin{eqnarray}
t(\phi) = \left(\frac{3}{V_{0_\phi}} \right)^{\frac{1}{2}} \frac{2
\left( 1 + \left( \frac{\beta_1}{\beta_2} \right)^{\frac{2}{m}}
\right)^{\frac{1}{2}}}{\beta_1^2 m^2} \phi^{\frac{\alpha_1}{2} - 2m + 2} \exp\left[
\frac{\beta_1 \phi^m}{2} \right].
\end{eqnarray}
Once more the evolution of $t(\phi)$ demonstrates that inflation
occurs at late times for large $\phi, \chi$. We then proceed in the same way and the solutions we get are as
follows:
\begin{eqnarray}
\phi(t) &=& \left( \frac{2}{\beta_1} \ln t \right)^{\frac{1}{m}},\\
\chi(t) &=& \left( \frac{2}{\beta_2} \ln t \right)^{\frac{1}{m}},\\
H(t) &=& \left( 1+ \left( \frac{\beta_1}{\beta_2}
\right)^{\frac{2}{m}} \right) \frac{1}{\beta_1 m^2} \left( \frac{2}{\beta_1}
\right)^{\frac{2 - m}{m}} \frac{1}{t}\left[ \ln t \right]^{\frac{2 -
2m}{m}},\\
a(t) &\sim& \exp\left[  \left( 1+ \left( \frac{\beta_1}{\beta_2}
\right)^{\frac{2}{m}} \right) \frac{1}{\beta_1 m (2-m)} \left[ \frac{2}{\beta_1} \ln t \right]^{\frac{2 - m}{m}} \right].
\end{eqnarray}
It should be noted that in the two exponential potential case this
reduces to power law inflation
\begin{eqnarray}
a(t) \sim t^{\frac{2(\beta_1^2 + \beta_2^2)}{\beta_1^2 \beta_2^2}},
\end{eqnarray}
which is agreement with the dynamical system analysis of \cite{coley}.
 The ``new'' feature is that we generate more
inflation than we would get with one scalar because of the extra factor, $\exp\left[  1+
\left(\frac{\beta_1}{\beta_2}\right)^{\frac{2}{m}} \right]$, in the
solution for the scale factor.
This concludes the analysis for the cases $m_1, m_2 > 0$.

\subsection{$m_1 \leq 0 , m_2 > 0$}
It is possible when $m_i < 0 $ and is also an even number, to generate a minimum of the
potential at the origin. The nature of the slow-roll parameters
dictates that this will certainly give a finite amount of inflation
and that we may not get slow-roll here. We shall assume, initially at
least, that the field begins in the part dominated by the power law part of
the potential. 

\subsubsection{$\alpha_1 \neq 0,-4 $}\label{sec:ref1}
Following the same procedure as the previous section we are able to
deduce that
\begin{eqnarray}
\dot{\phi} = \left( \frac{V_{0_\phi}}{3} \right)^{\frac{1}{2}}
\alpha_1 \phi^{-\frac{\alpha_1 + 2}{2}} \exp\left[ - \frac{\beta_1 
\phi^{m_1}}{2} \right].  \label{eq:phi5}
\end{eqnarray}
Taking the $\phi$ exponential as approximately constant, we are able
to solve this to give,
\begin{eqnarray}
t(\phi) = \left( \frac{3}{V_{0_\phi}} \right)^{\frac{1}{2}}
\frac{2}{\alpha_1(\alpha_1 + 4 ) } \phi^{\frac{\alpha_1+4}{2}} \exp\left[ \frac{\beta_1 
\phi^{m_1}}{2} \right].\label{eq:timeevo}
\end{eqnarray}
Solving for the remaining fields and parameters we deduce
\begin{eqnarray}
\phi(t)&=& \left( \left(\frac{ V_{0_\phi}}{3} \right)^{\frac{1}{2}}
\frac{\alpha_1}{2} (\alpha_1 + 4 ) t \right)^{\frac{2}{\alpha_1 + 4
}}, \\
\chi(t) &=& \left[ \frac{2}{\beta_2} \frac{ \alpha_1 + 2}{\alpha_1 +
4} \ln t \right]^{\frac{1}{m_2}},\\
H(t) &=& \left(\frac{ V_{0_\phi}}{3} \right)^{\frac{1}{2}}\left( \left(\frac{ V_{0_\phi}}{3} \right)^{\frac{1}{2}}
\frac{\alpha_1}{2} (\alpha_1 + 4 ) t
\right)^{-\frac{\alpha_1}{\alpha_1 + 4}}  \nonumber\\ && \times \exp\left[ -\frac{\beta_1}{2}\left( \left(\frac{ V_{0_\phi}}{3} \right)^{\frac{1}{2}}
\frac{\alpha_1}{2} (\alpha_1 + 4 ) t \right)^{\frac{2m_1}{\alpha_1 + 4
}}\right],\\
a(t) &\propto& \exp\left[ \frac{4+\alpha_1}{4}\left(\frac{ V_{0_\phi}}{3} \right)^{\frac{1}{2}}\left( \left(\frac{ V_{0_\phi}}{3} \right)^{\frac{1}{2}}
\frac{\alpha_1}{2} (\alpha_1 + 4 ) 
\right)^{-\frac{\alpha_1}{\alpha_1 + 4}} t^{\frac{4}{\alpha_1 + 4}} f_1(t) \right].
\end{eqnarray}
In this instance
\begin{eqnarray}
f_1(t) = \exp\left[ -\frac{\beta_1}{2}\left( \left(\frac{ V_{0_\phi}}{3} \right)^{\frac{1}{2}}
\frac{\alpha_1}{2} (\alpha_1 + 4 ) t \right)^{\frac{2m_1}{\alpha_1 + 4
}}\right],
\end{eqnarray}
and rapidly tends to unity as time increases. With this set of
parameters, studying (\ref{eq:timeevo}) one realizes that it is
possible for inflation to occur in a different manner. It is helpful
to write the scalefactor in the form 
\begin{eqnarray}
a(t) \propto \exp\left[ ct^k f_1(t) \right]
\end{eqnarray}
$ \alpha_1 > 0$: In this case, $\alpha_1(\alpha_1 + 4)
>0$ and so it follows, from equation \ref{eq:timeevo}, that $t(\phi) \rightarrow \infty$ as
$\phi$ increases. Therefore,  we must generate inflation at late
times.In addition,  $c>0, 0<k<1$ giving us a
relatively slow expansion rate.\\
$-4<\alpha_1 <0$: In this case  $\alpha_1(\alpha_1 + 4)
<0$ and we get $c>0$, $1<k<\infty$. As $\phi \rightarrow \infty$,
$t(\phi)\rightarrow -\infty$ and so inflation now occurs at early
times. Potentially this can also give us a lot of
inflation as the scale factor grows faster than $e^t$.\\
$\alpha_1 < 0 $: Once more we get $\alpha_1(\alpha_1 + 4)
>0$, however we now have a negative power of $\phi$ in
(\ref{eq:timeevo}) and so $t(\phi) \rightarrow 0$ as $\phi$
increases. This generates inflation at early times once more. In
addition we have the constraints $c,k >0$.

Once more the setup is equivalent to that of a single scalar as the
second scalar makes minimal contribution to the universal dynamics. We
generate identical results to section III.B of \cite{parsons}.

Let us now return to the case where $\chi$ dominates. This gives a
Hubble parameter of
\begin{eqnarray}
H(\phi) &=& \left(\frac{ V_{0_\phi} \alpha_1 (\alpha_1 +
2)}{3\beta_2^2 m_2^2}\right)^{\frac{1}{2}} \left( \frac{\alpha_1
+2}{\beta_2} \right)^{\frac{1-m_2}{m_2}} \phi^{-\frac{(\alpha_1 +
2)}{2}} \left[ \ln\phi\right]^{\frac{1}{m_2}-1} \exp\left[ -
\frac{\beta_1 \phi^{m_1}}{2} \right], \\
\dot{\phi} &=& \left(\frac{ V_{0_\phi} \alpha_1\beta_2^2 m_2^2}{3 (\alpha_1 +
2)}\right)^{\frac{1}{2}}\left( \frac{\alpha_1
+2}{\beta_2} \right)^{\frac{1-m_2}{m_2}} \phi^{-\frac{\alpha_1}{2}} \left[ \ln\phi\right]^{1-\frac{1}{m_2}} \exp\left[ -
\frac{\beta_1 \phi^{m_1}}{2} \right].
\end{eqnarray}
Integrating up, again regarding the exponential as almost constant we
get
\begin{eqnarray}
t(\phi) = \left( \frac{ 3 (\alpha_1 + 2)}{V_{0_\phi} \alpha_1
\beta_2^2}\right)^{\frac{1}{2}} \left( \frac{\alpha_1
+2}{\beta_2} \right)^{\frac{m_2-1}{m_2}}\phi^{\frac{\alpha_1 -2}{2}} \left[\ln
\phi\right]^{\frac{1}{m_2}} \exp\left[\frac{\beta_1 \phi^{m_1}}{2} \right].
\end{eqnarray}
We are not able to invert this equation in order to obtain a
meaningful description of $\phi(t)$. 

\subsubsection{$\alpha_1 = -4$}
The evolution equation for $\phi$ (\ref{eq:phi5}) still holds --
provided, of course, that we are still in a $\phi$ dominated regime. 
\begin{eqnarray}
\dot{\phi} = - 4 \left( \frac{V_{0_\phi}}{3} \right)^{\frac{1}{2}}
\phi \exp\left[ -\frac{\beta_1 \phi^{m_1}}{2} \right].
\end{eqnarray}
Then integrating we retrieve
\begin{eqnarray}
t(\phi) = -\frac{1}{4} \left( \frac{3}{V_{0_\phi}}
\right)^{\frac{1}{2}} (\ln \phi) \exp\left[\frac{\beta_1 \phi^{m_1}}{2}
\right]. \label{eq:timephi5}
\end{eqnarray}
Notice that we generate a $\ln$ rather than a power of $\phi$ in the
solution. This propagates through the calculation and will generate
qualitatively different behavior in the Hubble parameter and
scalefactor. However, we still see that $t(\phi) \rightarrow- \infty$
for increasing $\phi$ and so it is an early effect.
\begin{eqnarray}
\phi(t) &=& \exp\left[-4 \left( \frac{V_{0_\phi}}{3}
\right)^{\frac{1}{2}} t \right],\\
\chi(t) &=& \left[ \frac{8}{ \beta_2} \left( \frac{V_{0_\phi}}{3}
\right)^{\frac{1}{2}} t \right]^{\frac{1}{m_2}}, \\
H(t) &=& \left( \frac{V_{0_\phi}}{3}
\right)^{\frac{1}{2}} \exp\left[ -8\left( \frac{V_{0_\phi}}{3}
\right)^{\frac{1}{2}} t - \frac{\beta_1}{2} \exp\left[ -4 m_1 \left( \frac{V_{0_\phi}}{3}
\right)^{\frac{1}{2}} t \right] \right],\\
a(t) &\propto& \exp\left[ -\frac{1}{8} \exp\left[ -8 \left( \frac{V_{0_\phi}}{3}
\right)^{\frac{1}{2}} t \right] f_3(t) \right],
\end{eqnarray}
where 
\begin{eqnarray}
f_3(t) = \exp\left[ - \frac{\beta_1}{2} \exp\left[ -4 m_1\left( \frac{V_{0_\phi}}{3}
\right)^{\frac{1}{2}} t \right]\right].
\end{eqnarray} 
It is clear that $f_3(t) \rightarrow 1$ rapidly as $t$
increases. Looking at equation (\ref{eq:timephi5}), it is clear that
$t$ decreases for increasing $\phi$ so the inflation we generate is an
early time phenomenon.

\subsubsection{ $\alpha_1 = 0$}\label{sec:alpha0}
In this special case the $\phi$ field will quickly dominate once more,
as the potential for $\chi$ decays quickly. Thus we get behavior very
similar to that for one field, with $\chi$ just in the background as
in \cite{parsons}. Solving in the usual manner gives us
\begin{eqnarray}
\phi(t) &=& \left[\left( \frac{V_{0_\phi}}{3} \right)^{\frac{1}{2}}
\beta_1 m_1 ( 2 - m_1) \right]^{\frac{1}{2-m_1}}
t^{\frac{1}{2-m_1}},\\
\chi(t) &=& \left[ \frac{1}{\beta_2} \ln t \right]^{\frac{1}{m_2}},\\
H(t) &=& \left( \frac{V_{0_\phi}}{3} \right)^{\frac{1}{2}} \exp\left[
-\frac{\beta_1}{2} \left( \beta_1 m_1 (2-m_1) \left(
\frac{V_{0_\phi}}{3} \right)^{\frac{1}{2}} \right)^{\frac{m_1}{2-m_1}}
t^{\frac{m_1}{2-m_1}} \right], \\
a(t)&\propto& \exp\left[ \left( \frac{V_{0_\phi}}{3} \right)^{\frac{1}{2}} t
f_2(t)\right],
\end{eqnarray}
where
\begin{eqnarray}
f_2(t) =\exp\left[
-\frac{\beta_1}{2} \left( \beta_1 m_1 (2-m_1) \left(
\frac{V_{0_\phi}}{3} \right)^{\frac{1}{2}} \right)^{\frac{m_1}{2-m_1}}
t^{\frac{m_1}{2-m_1}} \right].
\end{eqnarray}
Inflation here occurs at early times.

\subsection{$ m_1, m_2 < 0$}
Proceeding in the usual manner we are able to show that
\begin{eqnarray}
H(\phi) &=& \left( \frac{V_{0_\phi}}{3} \right)^{\frac{1}{2}}
\phi^{-\frac{\alpha_1}{2}} \exp\left[ -\frac{\beta_1\phi^{m_1}}{2}
\right] \left( 1 + \frac{\alpha_1 (\alpha_1 + 2)}{\alpha_2 (\alpha_2 +
2)} \frac{\chi^2}{\phi^2} \right)^{\frac{1}{2}},\\
&=& \left( \frac{V_{0_\phi}}{3} \right)^{\frac{1}{2}}
\phi^{-\frac{\alpha_1}{2}} \exp\left[ -\frac{\beta_1\phi^{m_1}}{2}
\right] \nonumber \\
&&\times \left( 1 + \frac{\alpha_1 (\alpha_1 + 2)}{\alpha_2 (\alpha_2 +
2)} \left( \frac{ V_{0_\chi} \alpha_2 (\alpha_2 + 2) }{ V_{0_\phi}
\alpha_1 (\alpha_1 + 2)}\right)^{\frac{2}{\alpha_2 + 2}}\phi^{\frac{2(\alpha_1 - \alpha_2)}{\alpha_2 + 2 }}\right)^{\frac{1}{2}}.
\end{eqnarray}
It is then clear, for example, that $-2 < \alpha_1 < \alpha_2$ corresponds to $\phi$
domination. We shall consider this case and that of
equal contribution since $\chi$ domination is identical in behavior
to the former.
\subsubsection{$\alpha_2 > \alpha_1 > -2; \alpha_2 < -2 , \alpha_1>\alpha_2$} \label{sec:twoinverse}
If is clear that the $\phi$ field is dominant here and so the results
from section (\ref{sec:ref1}) hold for $\phi(t), H(t)$ and $a(t)$. In
this case we do get different evolution for $\chi(t)$ though, given by
\begin{eqnarray}
\chi(t) =\left( \frac{ V_{0_\chi} \alpha_2 (\alpha_2 + 2) }{ V_{0_\phi}
\alpha_1 (\alpha_1 + 2)}\right)^{\frac{1}{\alpha_2 + 2}} \left[ \left(\frac{V_{0_\phi}}{3} \right)
\frac{\alpha_1}{2}(\alpha_1 + 4) t \right]^{\frac{2(\alpha_1 +
2)}{(\alpha_1 + 4)(\alpha_2 + 2 )}}.
\end{eqnarray}
\subsubsection{$\alpha_1 = \alpha_2 = \alpha$}
This is the only case where we do not get domination by one field over
the other. Again the results will be very similar to that of section
(\ref{sec:ref1}) except that we will generate more inflation because
the Hubble parameter will be larger than before. Then we find that:
\begin{eqnarray}
t(\phi) &=& \left( \frac{3}{V_{0_\phi}} \right)^{\frac{1}{2}}
\frac{2\left( 1 + \left(\frac{V_{0_\chi}}{V_{0_\phi}} \right)^{\frac{2}{\alpha+2}}\right)^{\frac{1}{2}} }{\alpha(\alpha + 4)} \phi^{\frac{\alpha+4}{2}}
\exp \left[\frac{\beta_1 \phi^{m_1}}{2} \right] , \\
\phi (t) &=&\left(\left( \frac{V_{0_\phi}}{3}
\right)^{\frac{1}{2}} \frac{\alpha}{2}(\alpha+4)\left( 1 + \left(\frac{V_{0_\chi}}{V_{0_\phi}} \right)^{\frac{2}{\alpha+2}}\right)^{-\frac{1}{2}}  t
\right)^{\frac{2}{\alpha + 4}}, \\
\chi (t) &=&\left(
\frac{V_{0_\chi}}{V_{0_\phi}}\right)^{\frac{1}{\alpha + 2}} \left(\left( \frac{V_{0_\phi}}{3}
\right)^{\frac{1}{2}} \frac{\alpha}{2}(\alpha+4)\left( 1 + \left(\frac{V_{0_\chi}}{V_{0_\phi}} \right)^{\frac{2}{\alpha+2}}\right)^{-\frac{1}{2}}  t
\right)^{\frac{2}{\alpha + 4}}, \\
H(t) &=& \left( \frac{V_{0_\phi}}{3}
\right)^{\frac{1}{2}} \left( 1 + \left(\frac{V_{0_\chi}}{V_{0_\phi}}
\right)^{\frac{2}{\alpha+2}} \right)^{\frac{\alpha+2}{\alpha+4}}
\left(\left( \frac{V_{0_\phi}}{3}
\right)^{\frac{1}{2}} \frac{\alpha}{2}(\alpha+4)
t \right)^{-\frac{\alpha}{\alpha + 4}}\nonumber \\
&&\times \exp\left[ -\frac{\beta_1}{2}\left(\left( \frac{V_{0_\phi}}{3}
\right)^{\frac{1}{2}} \frac{\alpha}{2}(\alpha+4)\left( 1 + \left(\frac{V_{0_\chi}}{V_{0_\phi}} \right)^{\frac{2}{\alpha+2}}\right)^{-\frac{1}{2}}  t\right)^{\frac{2m_1}{\alpha + 4}}\right], \\
a(t) &\propto& \exp\left[ \frac{4+\alpha}{4}  \left( \frac{V_{0_\phi}}{3}
\right)^{\frac{1}{2}} \left( 1 + \left(\frac{V_{0_\chi}}{V_{0_\phi}} \right)^{\frac{2}{\alpha+2}} \right)^{\frac{\alpha+2}{\alpha+4}}
\left(\left( \frac{V_{0_\phi}}{3}
\right)^{\frac{1}{2}} \frac{\alpha}{2}(\alpha+4)
\right)^{-\frac{\alpha}{\alpha + 4}} t^{\frac{4}{\alpha +4}} f_4(t)\right].
\end{eqnarray}
where
\begin{eqnarray}
f_4(t) = \exp\left[ -\frac{\beta_1}{2}\left(\left( \frac{V_{0_\phi}}{3}
\right)^{\frac{1}{2}} \frac{\alpha}{2}(\alpha+4)\left( 1 + \left(\frac{V_{0_\chi}}{V_{0_\phi}} \right)^{\frac{2}{\alpha+2}}\right)^{-\frac{1}{2}}  t\right)^{\frac{2m_1}{\alpha + 4}}\right].
\end{eqnarray}
$f_4(t) \rightarrow 1$ rapidly as $t$ increases. This gives us
inflation at late times once more.\\

\vspace{0.5cm}
\begin{center}
\textbf{Small Field Values}
\end{center}

\vspace{0.5cm}

A moments thought reveals that in order to generate slow--roll at small
values of either $\chi$ or $\phi$ we must have $\alpha_i = 0, m\geq
1$. This can be seen easily by examining the slow--roll parameters
(\ref{eq:slowparam}). Our analysis will then follow a similar pattern
to before. Again,  we shall see
that one field will become dominant if $m_1 \neq m_2$ and we proceed
as before. 
\subsection{$m_1 < m_2, m_1 \neq2$}
Once more the $\phi$ will provide the dominant contribution and so
\begin{eqnarray}
H(\phi) = \left(\frac{V_{0_\phi}}{3}\right)^{\frac{1}{2}}\exp\left[-\frac{\beta_1\phi^{m_1}}{2}\right].
\end{eqnarray}
Then we find that
\begin{eqnarray}
\dot{\phi} =  \left(\frac{V_{0_\phi}}{3}\right)^{\frac{1}{2}} \beta m_1
\phi^{m_1 -1} \exp\left[-\frac{\beta_1\phi^{m_1}}{2}\right].\label{eq:phismall} 
\end{eqnarray}
This is identical to the case we considered previously in section
(\ref{sec:alpha0}) and so all the results there still hold. The one
difference is that the evolution of $\chi$ changes slightly.
\begin{eqnarray}
\chi(t) = \left[ \left(\frac{V_{0_\phi}}{3}\right)^{\frac{1}{2}}
\frac{V_{0\phi}}{V_{0\chi}} \beta_2 m_2 ( 2-m_2) t
\right]^{\frac{1}{m_2 - 2}}.\
\end{eqnarray}
One should
note that as $\phi\rightarrow 0$, $t$ decreases and so we are looking
at an early time effect.  This is equivalent to section IV.A of
\cite{parsons}. The cases where either one or both of $m_i = 2$ are easily covered and
does not result in any analytical difficulties. The behavior is not
markedly different from the above.

\subsection{$2 = m_1 < m_2$}
The $\phi$ is still dominant and so we can integrate
(\ref{eq:phismall}) to achieve
\begin{eqnarray}
t(\phi) = \left(\frac{3}{V_0}\right)^{\frac{1}{2}} \frac{1}{2\beta_1}
(\ln \phi) \exp\left[-\frac{\beta_1\phi^2}{2}\right].
\end{eqnarray}
Solving the equation for small $\phi$ we find that
\begin{eqnarray}
\phi(t) &=& \exp\left[2\left(\frac{V_{0_\phi}}{3}\right)^{\frac{1}{2}} \beta
t\right],\\
\chi(t) &=& \left[ 4\left(\frac{V_{0\phi}}{3}\right)^{\frac{1}{2}}
\frac{V_{0_\phi}}{V_{0_\chi}} \frac{\beta_1^2}{\beta_2 m_2 ( 2-m_2) }
t \right]^{\frac{1}{m_2 -2}}.
\end{eqnarray}
Note that although the behavior $\chi(t) \sim (-\kappa_1 t
)^{\kappa_2}, \kappa_1 > 0$ appears problematic, this would be
resolved by including the integration constants. This would then give
us 
\begin{eqnarray}
\chi(t) &=& \left[ K - 4\left(\frac{V_{0\phi}}{3}\right)^{\frac{1}{2}}
\frac{V_{0_\phi}}{V_{0_\chi}} \frac{\beta_1^2}{\beta_2 m_2 ( 2-m_2) }
t \right]^{\frac{1}{m_2 -2}}, \ \ \ K > 0.
\end{eqnarray}
Now since $\phi(t)$ in a monotonic increasing function, the slow--roll
conditions will be violated within finite time and so our solution
will no longer be valid. This should occur before $ t =  \frac{K}{4}  \left(\frac{3}{V_{0\phi}}\right)^{\frac{1}{2}}
\frac{V_{0_\chi}}{V_{0_\phi}} \frac{\beta_2 m_2 ( 2-m_2) }{\beta_1^2}$
since there are no physical grounds upon which to rule out this solution.

\subsection{$m_1 = m_2 = 2$}
In this instance we find that
\begin{eqnarray}
H(\phi) = \frac{1}{\sqrt{3}} \left( V_{0_\phi} +  V_{0_\chi}
\right)^{\frac{1}{2}} \exp\left[ -\frac{\beta_1 \phi^2}{2}\right].
\end{eqnarray}
It is not too difficult to show that
\begin{eqnarray}
t(\phi) = \frac{\sqrt{3}}{2\beta_1 V_{0_\phi}} \left( V_{0_\phi} +  V_{0_\chi}
\right)^{\frac{1}{2}}(\ln \phi) \exp\left[\frac{\beta_1 \phi^2}{2}
\right].
\end{eqnarray}
It is clear that as $\phi \rightarrow 0$ we have $t \rightarrow
-\infty$ and so inflation will occur at early times. Then, proceeding
in the familiar manner, we can show
\begin{eqnarray}
\phi(t) &=& \exp\left[ \frac{V_{0_\phi}}{\sqrt{3}} \frac{ 2\beta_1}{\left( V_{0_\phi} +  V_{0_\chi}
\right)^{\frac{1}{2}}} t \right],\\
\chi(t) &=& \exp\left[ \frac{V_{0_\chi}}{\sqrt{3}} \frac{ 2\beta_2}{\left( V_{0_\phi} +  V_{0_\chi}
\right)^{\frac{1}{2}}} t \right],\\
H(t) &=& \frac{1}{\sqrt{3}} \left( V_{0_\phi} +  V_{0_\chi}
\right)^{\frac{1}{2}} \exp\left[ -\frac{\beta_1}{2} \exp\left[ \frac{V_{0_\phi}}{\sqrt{3}} \frac{ 2\beta_1}{\left( V_{0_\phi} +  V_{0_\chi}
\right)^{\frac{1}{2}}} t \right] \right],\\
a(t) &\propto& \exp\left[ H(t) t \right].
\end{eqnarray}
where we have treated all the exponentials as being approximately
equal to unity.

\vspace{0.5cm}
\begin{center}
\textbf{Large and small field values}
\end{center}

\vspace{0.5cm}

If we take one of the fields large and one small we will always have
the large field dominant and so this relates to the cases considered
already. For example, small $\chi$ means that $m_2 \geq 1$ and so $m_1
< m_2$. Therefore the $\phi$ field will be dominant. This applies to
all possible cases.

\subsection{Summary for Uncoupled Potential}
We have seen that for  two fields with uncoupled potentials often
one field will be dominant, and we return to an effective single field theory
-- for the background at least-- as studied in
\cite{parsons}. There are, however,  several cases where the
second field is important and we have demonstrated them. We
have seen that it is possible to generate inflation at both late and
early times and for large and small values of both of the fields. As a
result, these background solutions do not differ qualitatively from
the single field set-up \cite{parsons}. The
results are summarized in the table below. If $m_1 < m_2$ then the
$\phi$ field is dominant so we only outline the results for $m_1$. If
 $m_1 = m_2 = m$ the same qualitative behavior exists except that we
generate more inflation than we would normally with one field.

\begin{center}
\begin{table}[!h]
\caption{Summary of Inflationary Behavior for uncoupled potentials} \label{tab:uncoup}
\begin{tabular}{c|c|c}
Early Times & Late Times & All Times\\
\hline
$m_1 \leq 0, \alpha_1 \leq 0$ & $0\leq m_1 \leq 1$ & $m_1 = 1,
\alpha_1 = 0$ \\
or & or & or \\
$m_1 \geq 1, \alpha_1 = 0$ & $m_1 \leq 0, \alpha_1 \geq 0 $ & $m_1 =
\alpha_1 = 0$ \\
\end{tabular}
\end{table}
\end{center}

\section{A coupled potential}
We now turn our attention to a potential of the form 
\begin{eqnarray}\label{coupledpot}
V(\phi,\chi)  = V_0 \phi^{-\alpha} e^{-\beta\chi^m}
\end{eqnarray}
and repeat the processes of the previous section. The motivation for
such a potential arises, for example, in higher--dimensional theories
of the form
\begin{equation}
S = \int d^N x \sqrt{-g_N}\left[ \frac{1}{2\kappa_N}R - 
\frac{1}{2}(\partial \phi)^2 - V(\phi) \right].
\end{equation}
Writing the $N$--dimensional metric in the form
\begin{equation}
ds^2_N = g^{(4)}_{\mu\nu}dx^\mu dx^\nu + b^2 ds_D^2,
\end{equation}
and compactifying the theory on a torus which has zero curvature, one 
ends up with a potential of the form (\ref{coupledpot}), where 
$\chi \propto \ln b$ describing the size of the extra dimensions 
(see e.g. \cite{starobinsky}). 

We still have the three coupled equations
(\ref{eq:slow1}-\ref{eq:slow3}) to solve which are equivalent to
\begin{eqnarray}
H &=& \left(\frac{V_0}{3}\right)^{\frac{1}{2}} \phi^{-\frac{\alpha}{2}}\exp\left[-\frac{\beta\chi^m}{2}\right],\label{eq:hubcoup}\\
3H\dot{\phi} &=& \alpha V_0\phi^{-(\alpha+1)} \exp\left[-\beta\chi^m\right], \\
3H\dot{\chi} &=& \beta m \chi^{m-1}V_0\phi^{-\alpha} \exp\left[-\beta\chi^m\right],
\end{eqnarray}
with solution
\begin{eqnarray}
\phi^2 = \left\{\begin{array}{cc} \frac{2\alpha}{\beta m ( 2- m)}
\chi^{2-m} , &
\ \ \ m \neq 2,  \label{eq:couprel}\\
\frac{\alpha}{\beta} \ln \chi,& \ \ \ m = 2.
\end{array}\right.
\end{eqnarray}
There are certain cases where it is not clear that equation
(\ref{eq:couprel}) means anything-- for example $m>2 , \alpha > 0$-- since this would generate an imaginary value for at least one
of the fields. At this stage one should remember that we have omitted
all integration constants up to this point. As there are no physical
grounds on which to rule out such cases, when one puts back these
constants (dependent on the initial conditions), any such problems may
be overcome.

Before we begin, we shall find it instructive to examine some plots of
the potential in question. These are shown in figure 1 
where
we have plotted $V(\phi, \chi ) = \phi^{-\frac{1}{2}} \exp\left[-2\chi^{\pm
0.2}\right]$ respectively.

\begin{figure}[!ht] \label{plots}
\centerline{\includegraphics{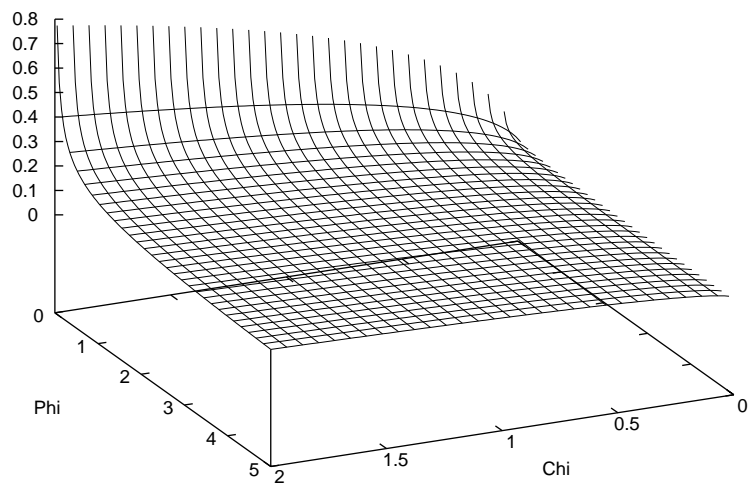}}
\centerline{\includegraphics{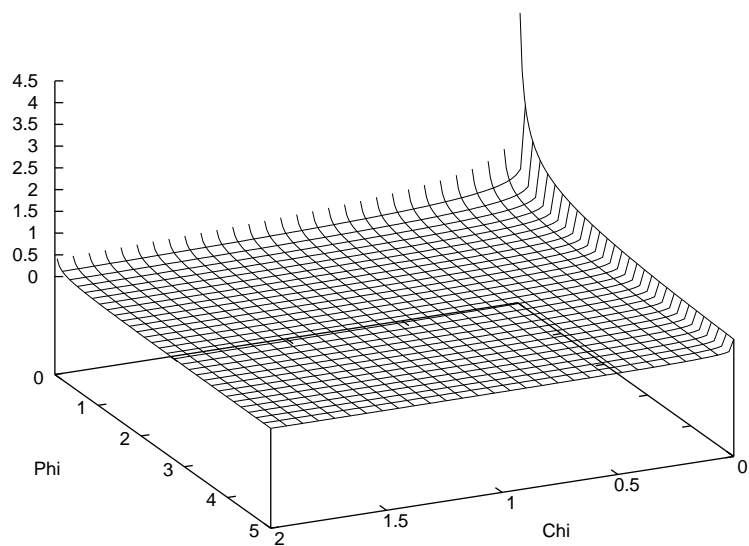}}
\caption{Plot of the Coupled potential.}
\end{figure}


\begin{center}
\textbf{Large field Values}
\end{center}

\subsection{$1 \geq m>0$}\label{sec:m<1}
This guarantees us slow--roll for large values of the fields.
We generate the following equation for $\chi$
\begin{eqnarray}
\dot{\chi} = \left( \frac{V_0}{3}\right)^{\frac{1}{2}} \beta m
\left( \frac{\beta m (2-m)}{2\alpha}\right)^{\frac{\alpha}{4}} \chi^{\frac{\alpha(m-2)}{4}} \chi^{m-1} \exp\left[\frac{-\beta
\chi^m}{2}\right].\label{eq:chidotcoup}
\end{eqnarray}
We are then able to solve this to produce the following expression,
\begin{eqnarray}
t(\chi) = \left( \frac{3}{V_0}\right)^{\frac{1}{2}} \frac{2}{
\beta^2 m^2} \left( \frac{\beta m
(2-m)}{2\alpha}\right)^{-\frac{\alpha}{4}} \chi^{\frac{\alpha(2-m)}{4}}
\chi^{2-2m} \exp\left[\frac{\beta
\chi^m}{2}\right].\label{eq:tcoup1}
\end{eqnarray}
Proceeding in a similar, way we are able to solve the system producing,
to leading order in $t$, the results below.
\begin{eqnarray}
\phi(t) &=& \left( \frac{2\alpha}{\beta m (2-m)} \right)^{\frac{1}{2}} \left[
\frac{2}{\beta} \ln t \right]^{\frac{2-m}{2m}},\label{eq:phicoup2} \\
\chi(t) &=& \left[ \frac{2}{\beta} \ln t \right]^{\frac{1}{m}}, \\
H(t) &=&\frac{2}{\beta^2 m^2} \frac{1}{t} \left[ \frac{2}{\beta}
\ln t \right]^{\frac{2-2m}{m}}, \label{eq:hubbcoup}\\
a(t) &\propto& \exp\left[ \frac{1}{\beta m (2-m)}
\left[\frac{2}{\beta}\ln t\right]^{\frac{2-m}{m}} \right].  
\end{eqnarray}

One should note that this gives very similar behavior to section
(\ref{sec:firstuncoup}). This should not be entirely unexpected
because the exponential part of the potential will always dominate for $m>0$.

\subsection{$m<0$} \label{sec:coup2}
Due to the nature of the slow--roll
parameters, we must once more consider large values for both of the
fields. Now solving for $\phi$, treating the exponential approximately
constant, we find 
\begin{eqnarray}
t(\phi, \chi) =  \left( \frac{3}{V_0}\right)^{\frac{1}{2}}
\frac{2}{\alpha(\alpha+4)} \phi^{\frac{\alpha +4}{2}}  \exp\left[\frac{\beta\chi^m}{2}\right].
\end{eqnarray}
Substituting for either field using~(\ref{eq:couprel}) we able to invert this finding
\begin{eqnarray}
\chi(t) &=& \left( \frac{\beta m (2-m)}{2\alpha}
\right)^{\frac{2}{2-m}}\left[ \left(\frac{V_0}{3}\right)^{\frac{1}{2}}
\frac{\alpha}{2} (\alpha + 4) t \right]^{\frac{4}{(2-m)(\alpha + 4)}},\\
\phi(t) &=& \left[\left(\frac{V_0}{3}\right)^{\frac{1}{2}}
\frac{\alpha}{2} (\alpha + 4) t \right]^{\frac{2}{\alpha + 4}}.
\end{eqnarray}
One should also note that the onset of inflation occurs in analogy
with section (\ref{sec:ref1}). It is also worth
pointing out that this gives the solution one would expect when
$m\rightarrow 0$. That is, we generate the same results as
(\ref{sec:ref1}) except that the $\chi$ field freezes out. This should
not surprise us since the potential becomes equivalent to that of an
inverse power.  We then find the form for the Hubble parameter 
\begin{eqnarray}
H(t) &=& \left(\frac{V_0}{3}\right)^{\frac{1}{2}}\left[\left(\frac{V_0}{3}\right)^{\frac{1}{2}}
\frac{\alpha}{2} (\alpha + 4) t \right]^{-\frac{\alpha}{\alpha +
4}}\nonumber\\
&&\times \exp\left[- \frac{\beta}{2}  \left( \frac{\beta m (2-m)}{2\alpha}
\right)^{\frac{2m}{2-m}}\left[ \left(\frac{V_0}{3}\right)^{\frac{1}{2}}
\frac{\alpha}{2} (\alpha + 4) t \right]^{\frac{4m}{(2-m)(\alpha + 4)}}\right],
\end{eqnarray}
where, of course, the exponential part tends rapidly to
1. Solving for the scale factor we deduce to leading order
\begin{eqnarray}
a(t) \propto  \exp\left[ \frac{4+\alpha}{4}\left(\frac{
V_0}{3} \right)^{\frac{1}{2}}\left( \left(\frac{ V_0}{3}
\right)^{\frac{1}{2}} \frac{\alpha}{2} (\alpha + 4 ) 
\right)^{-\frac{\alpha}{\alpha + 4}} t^{\frac{4}{\alpha + 4}}
g_1(t) \right],
\end{eqnarray}
where 
\begin{eqnarray}
g_1(t) = \exp\left[- \frac{\beta}{2}  \left( \frac{\beta m (2-m)}{2\alpha}
\right)^{\frac{2m}{2-m}}\left[ \left(\frac{V_0}{3}\right)^{\frac{1}{2}}
\frac{\alpha}{2} (\alpha + 4) t \right]^{\frac{4m}{(2-m)(\alpha + 4)}}\right].
\end{eqnarray}

It is obvious that the special case of $\alpha = 0$ gives us just a
single scalar field that evolves. This has already been studied. We
must however consider what happens when $\alpha = -4$. We generate the
following equation
\begin{eqnarray}
t(\phi, \chi) = -\frac{1}{4} \left(\frac{3}{V_0} \right)^{\frac{1}{2}}
[\ln \phi] \exp\left[-\frac{\beta\chi^m}{2} \right].
\end{eqnarray}
Then solving for the fields we find
\begin{eqnarray}
\phi(t) &=& \exp\left[ -4 \left( \frac{V_0}{3} \right)^{\frac{1}{2}} t
\right], \\
\chi(t) &=& \exp\left[ -\frac{8}{2-m} \left( \frac{V_0}{3} \right)^{\frac{1}{2}} t
\right], \\
H(t) &=&  \left( \frac{V_0}{3} \right)^{\frac{1}{2}} \exp\left[ -8
\left( \frac{V_0}{3} \right)^{\frac{1}{2}} t - \frac{\beta}{2}  \exp\left[ -\frac{8m}{2-m} \left( \frac{V_0}{3} \right)^{\frac{1}{2}} t
\right]\right], \\
a(t) &\propto&\exp\left[ - \frac{1}{8} \exp\left[ -8
\left( \frac{V_0}{3} \right)^{\frac{1}{2}} t \right] g_2(t) \right]
\end{eqnarray}
where 
\begin{eqnarray}
g_2(t) =  \exp\left[ - \frac{\beta}{2}  \exp\left[ -\frac{8m}{2-m} \left( \frac{V_0}{3} \right)^{\frac{1}{2}} t
\right]\right],
\end{eqnarray}
which rapidly tends to unity.
\subsection{$1<m<2$}

If one considers the slow--roll parameters we may expect only to generate
inflation for small values of $\chi$. However, we must still take
large $\phi$ for slow--roll to be a valid approximation. In this
instance, we are able to approximate the exponential by unity. Provided
$m \neq 2$, we  generate the same set of results as for $m<0$ with
large $\chi$, as in section \ref{sec:coup2}.

In the special case where $m = 2$, we get slightly different
behavior. We summarize the results below:
\begin{eqnarray}
\phi(t) &=& \left[\left(\frac{V_0}{3}\right)^{\frac{1}{2}}
\frac{\alpha}{2} (\alpha + 4) t \right]^{\frac{2}{\alpha + 4}},\\
\chi(t) &=& \exp\left[ \frac{\beta}{\alpha}\left[\left(\frac{V_0}{3}\right)^{\frac{1}{2}}
\frac{\alpha}{2} (\alpha + 4) t \right]^{\frac{2}{\alpha + 4}}\right],
\\
H(t) &=& \left(\frac{V_0}{3}\right)^{\frac{1}{2}}\left[\left(\frac{V_0}{3}\right)^{\frac{1}{2}}
\frac{\alpha}{2} (\alpha + 4) t \right]^{-\frac{\alpha}{\alpha +
4}} \exp\left[ -\frac{\beta}{2} \exp\left[ \frac{m\beta}{\alpha}\left[\left(\frac{V_0}{3}\right)^{\frac{1}{2}}
\frac{\alpha}{2} (\alpha + 4) t \right]^{\frac{2}{\alpha +
4}}\right]\right],\\
a(t) &\propto& \exp\left[ \frac{4+\alpha}{4}\left(\frac{
V_{0_\phi}}{3} \right)^{\frac{1}{2}}\left( \left(\frac{ V_{0_\phi}}{3}
\right)^{\frac{1}{2}} \frac{\alpha}{2} (\alpha + 4 ) 
\right)^{-\frac{\alpha}{\alpha + 4}} t^{\frac{4}{\alpha + 4}}
g_3(t) \right],
\end{eqnarray}
where
\begin{eqnarray}
g_3(t) =  \exp\left[ -\frac{\beta}{2} \exp\left[ \frac{m\beta}{\alpha}\left[\left(\frac{V_0}{3}\right)^{\frac{1}{2}}
\frac{\alpha}{2} (\alpha + 4) t \right]^{\frac{2}{\alpha +
4}}\right]\right].
\end{eqnarray}

\subsection{Summary for the Coupled Potential}

The summary of the inflationary behavior is effectively equivalent to
Table \ref{tab:uncoup} with the obvious changes to the parameters. We
are able to generate inflation at late, early or all times  by tuning the
parameters in the potential. If $m>0$, the exponential part is
dominant and the solutions produced are similar to that of a single
scalar, $\chi$ with the second field, $\phi$, in the background. When
$m\leq0$, the exponential part is subdominant and the results resemble
those for a single scalar, $\phi$, with a $\phi^{-\alpha}$
potential. The quantitative modifications are small and quickly
decrease with time. 

In all cases one of the fields provides the dominant contribution to
$H(t)$ and $a(t)$. The effects of the second field are
present but decay quickly so that the field is only in the background.

\section{Cosmological Perturbations}
We now turn our attention to the evolution for cosmological
perturbations in both of the scalar fields and the spacetime 
metric during a period of inflation. 
Again, we assume that the fields are slow--rolling, implying that the
parameters (\ref{eq:slowparam}) are small.  Furthermore, we shall
assume that both fields contribute approximately equally to the
expansion rate. Only then can we expect new features in the
perturbation spectra resulting from the second field.

A convenient formalism to 
study both adiabatic and isocurvature perturbations in inflation 
was presented in \cite{gordon}, which we will use here. Instead of 
working with the fields $\phi$ and $\chi$ it is useful to 
perform a rotation as follows:
\begin{eqnarray}
\delta \sigma &=& (\cos \theta) \delta\phi   +   (\sin \theta) \delta\chi \\
\delta s &=& -  (\sin \theta) \delta\phi + (\cos \theta) \delta\chi,
\end{eqnarray}
with
\begin{equation}
\cos \theta = \frac{\dot\phi}{\sqrt{\dot\phi^2 + \dot\chi^2}}, \hspace{1cm}
\sin \theta = \frac{\dot\chi}{\sqrt{\dot\phi^2 + \dot\chi^2}}.
\end{equation}
$\sigma$ is called the adiabatic field and $s$ is called the 
entropy field. The motivation for their names becomes 
clear when one considers fluctuations of them. 

The line--element for arbitrary scalar perturbations of the
Robertson--Walker metric for a spatially flat universe 
reads, (using the notation of \cite{gordon})
\begin{eqnarray}
ds^2 &=& -(1+2A)dt^2 + 2 a^2 B_{,i} dx^i dt \nonumber \\
&+& a^2 \left[(1-2\psi)\delta_{ij} + 2 E_{,ij}\right]dx^i dx^j.
\end{eqnarray}
The gauge--invariant curvature perturbation, defined as 
\begin{equation}
{\cal R} = \psi + \frac{H \delta \rho}{\dot \rho}, \label{curvpert}
\end{equation}
is, on very large scales, constant for purely adiabatic
perturbations (see e.g. \cite{Malik} and references therein). 
However, entropy perturbations are a source for the curvature perturbation
(\ref{curvpert}). The entropy perturbation between two 
species $A$ and $B$, defined as 
\begin{equation}
{\cal S} = \frac{\delta n_A}{n_A} - \frac{\delta n_B}{n_B},
\end{equation}
where $n_i$ are the number densities of the particle species $i$ can 
evolve in time, even on superhorizon scales. Therefore, it was argued
\cite{wands} that on very large scales in general we have the following equations
describing the evolution of ${\cal R}$ and 
${\cal S}$:
\begin{eqnarray}
\dot {\cal R} &=& \gamma H {\cal S}, \label{eomR}\\
\dot {\cal S} &=& \delta H {\cal S}. \label{eomS}
\end{eqnarray}
For the case of the two slow--rolling scalar fields and in the spatial
flat gauge ($\psi=0$), ${\cal R}$ and ${\cal S}$ are given
by
\begin{equation}
{\cal R} \approx \frac{H\left( \dot\phi \delta\phi + \dot\chi
\delta\chi\right)}{\dot\phi^2 + \dot\chi^2} = \frac{H \delta
\sigma}{\dot \sigma}
\end{equation}
and 
\begin{equation}
{\cal S} = \frac{H\left( \dot\phi \delta\chi - \dot\chi
\delta\phi\right)}{\dot\phi^2 + \dot\chi^2} = \frac{H \delta s}{\dot \sigma}.
\end{equation}
For ${\cal S}$ we use the same normalization as in
\cite{wands}. Fluctuations in the field $\sigma$ are adiabatic perturbations, 
whereas fluctuations in $s$ are entropic perturbations. On very large
scales ($k\ll aH$) and in flat gauge,
the evolution of fluctuations are described as \cite{gordon}
\begin{eqnarray}
(\delta \sigma)^{..} + 3 H (\delta \sigma)^. + 
\left( V_{\sigma\sigma} - \dot\theta^2\right)\delta\sigma =
-2 V_\sigma A + \dot\sigma \dot A + 2(\dot\theta\delta s)^. - 
2\frac{V_\sigma}{\dot \sigma}\dot \theta \delta s \label{adiafield} 
\end{eqnarray}
and
\begin{eqnarray}
(\delta s)^{..} + 3 H (\delta s)^. + 
\left( V_{ss} - \dot\theta^2\right)\delta s  = 
-2\frac{\dot\theta}{\dot\sigma}\left[ \dot\sigma((\delta \sigma)^. -
\dot\sigma A) -\ddot \sigma \delta\sigma\right].\label{entrofield}
\end{eqnarray} 
The metric perturbation $A$ can be obtained from Einstein's equation 
and is given, in the flat gauge, by
\begin{equation}
H A =  4\pi G \left(\dot\phi \delta\phi + \dot\chi \delta\chi\right).
\end{equation}
An important point is that $\dot\theta$ must be non--vanishing in order 
for the adiabatic field to be sourced by the entropy field. If 
$\theta$ is constant, the entropy field does not contribute to 
perturbations in the gravitational potential.

In \cite{wands} the formalism presented was applied to slow--roll
inflation with two scalar fields and it was shown that 
$\gamma$ and $\delta$ in (\ref{eomR}) and (\ref{eomS}) 
are given in terms of the slow--roll parameters
\begin{eqnarray}
\gamma &=& -2\eta_{\sigma s}, \label{alpha} \\
\delta &=& -2\epsilon + \eta_{\sigma\sigma} - \eta_{ss} \label{beta},
\end{eqnarray}
and are, therefore, specified by the potential $V(\phi,\chi)$. In the 
last two equations, the slow--roll parameters are constructed from the 
usual slow--roll parameters for $\phi$ and $\chi$ (\ref{eq:slowparam}) and are given by
\begin{equation}\label{epsilongr}
\epsilon = \frac{1}{2}\left(\frac{V_{\sigma}}{V}\right)^2 
\approx \epsilon_\phi + \epsilon_\chi. 
\end{equation}
and
\begin{eqnarray}
\eta_{\sigma\sigma} &=& \eta_{\phi\phi}\cos^2\theta + 
2\eta_{\phi\chi}\cos\theta\sin\theta + 
\eta_{\chi\chi} \sin^2 \theta, \nonumber \\
\eta_{s s} &=& \eta_{\phi\phi}\sin^2\theta -  
2\eta_{\phi\chi}\cos\theta\sin\theta + 
\eta_{\chi\chi} \cos^2 \theta,\label{etagr} \\
\eta_{\sigma s} &=&
(\eta_{\chi\chi}-\eta_{\phi\phi})\sin\theta\cos\theta 
+ \eta_{\phi\chi}(\cos^2 \theta - \sin^2 \theta). \nonumber
\end{eqnarray}
The time evolution of ${\cal R}$ and ${\cal S}$ between horizon
crossing and some later time is given by
\begin{eqnarray}
\left(\begin{array}{c}
{\cal R} \\
{\cal S} \end{array} \right) = 
\left( \begin{array}{cc} 
1 & T_{{\cal R}{\cal S}} \\
0 & T_{{\cal S}{\cal S}} \end{array} \right) \left(\begin{array}{c}
{\cal R} \\
{\cal S} \end{array} \right)_*, \label{eq:matrixevo}
\end{eqnarray}
where the asterisk marks the time of horizon crossing. The transfer
functions $T_{{\cal R}{\cal S}}$ and $T_{{\cal S}{\cal S}}$ are given
by 
\begin{eqnarray} 
T_{{\cal S}{\cal S}}(t,t_*) &=& \exp\left( \int_{t_*}^t \delta(t') 
H(t') dt' \right)  \nonumber \\
T_{{\cal R}{\cal S}}(t,t_*) &=& \int_{t_*}^t \gamma(t')T_{{\cal
S}{\cal S}}(t_*,t') H(t') dt'. \label{transfer}
\end{eqnarray}
Thus, in the case of two slow--rolling scalar fields, the transfer
functions are completely specified by the potential through the 
slow--roll parameters. 

We are now in a position to calculate some of these transfer functions
for our general potentials. We shall consider the most simple
cases only in order to show how the background solutions derived 
in Section 3 and Section 4 can be used in order to study
perturbations. 

\subsection{Analytic Calculations}
We begin with analytic calculations for two simple cases. 
The first case considers two exponential potentials (a special 
case of assisted inflation). The second case is a simple model 
of a coupled potential. 
\subsubsection{Two Exponentials}
Let us now take
\begin{eqnarray}
V(\phi,\chi) = V_{0_\phi} e^{-\beta_1 \phi} +  V_{0_\chi} e^{-\beta_2
\chi}.
\end{eqnarray}
Although this setup was studied earlier \cite{assisted}, we would 
like to check the formalism for the perturbations. We return to the
first approximation (\ref{eq:firstapprox}) to make the
calculation. Then in this instance we find
\begin{eqnarray}
V_{0_\phi} e^{-\beta_1 \phi} &=& \frac{\beta_2^2}{\beta_1^2} V_{0_\chi} e^{-\beta_2
\chi}, \\
\dot{\phi} &=& \frac{\beta_2}{\beta_1} \dot{\chi}.
\end{eqnarray} 
These two equations enable us to calculate all the parameters. We find that
\begin{eqnarray}
\cos \theta = \frac{\beta_2}{\sqrt{\beta_1^2 + \beta_2^2}},\ \ \ \sin
\theta =\frac{\beta_1}{\sqrt{\beta_1^2 + \beta_2^2}}.
\end{eqnarray}
This shows that we get a constant angle in the phase plane. A quick
glance at equations (\ref{adiafield}-\ref{entrofield}) reveals that
the adiabatic and entropy perturbations decouple in this case. We
should expect our analysis to reflect this.
It is also straightforward to show that
\begin{eqnarray}
2 \epsilon = \eta_{\sigma\sigma} &=& \eta_{ss} = \frac{\beta_1^2 \beta_2^2}{\beta_1^2 +
\beta_2^2},\\
\eta_{\sigma s} &=& 0.
\end{eqnarray}
The functions $\gamma(t), \delta(t)$ then turn out to be constant given
by
\begin{eqnarray}
\gamma &=& 0, \\
\delta &=& - \frac{\beta_1^2 \beta_2^2}{\beta_1^2 +
\beta_2^2} <0.
\end{eqnarray}
Then we simply compute the transfer functions using (\ref{transfer}) to see that
\begin{eqnarray}
T_{\cal{SS}} (t,t_*) &=& \left( \frac{a(t)}{a(t_*)} \right)^\delta, \\
T_{\cal{RS}}  (t,t_*) &=& 0.
\end{eqnarray}
Now since $\delta<0$, it is immediately obvious that $T_{\cal{SS}}
(t,t_*) \rightarrow 0$ as time passes, provided $t>t_*$. Secondly, 
$T_{\cal{RS}} (t,t_*) =0$. This means that $\cal{R}$ remains constant and $\cal{S}$ $ \rightarrow 0 $ from (\ref{eq:matrixevo}).
Therefore, in this case we only expect adiabatic perturbations and no 
entropy perturbations if inflation lasts long enough. This is in
agreement with the results in \cite{assisted}. One should add that
the transfer functions are calculated from the time of horizon
crossing. However, when we calculate the transfer parameters,
$\gamma(t), \delta(t)$, we assume that $k/aH \ll 1$. This is only true
a few e--folds after horizon crossing. This means that our calculated
transfer functions will be accurate a few e--folds after horizon
crossing and so, in effect, they have the wrong initial
conditions. Whilst this does not affect their predictions, this means
they differ from what we observe numerically. In this instance there
is some evolution of $\mathcal{R}$ before it freezes in and some evolution
of $\mathcal{S}$ before if follows the decay predicted. This can be
seen in Figure \ref{fig:expspec} in the first few e--folds after
horizon exit.

\subsubsection{ A coupled example}
The next simplest case will prove to be that with a
coupled potential with a straight exponential. That is
\begin{eqnarray}
V(\phi,\chi) = V_0 \phi^{-\alpha} e^{-\beta \chi}.
\end{eqnarray}
We shall find it most convenient to work in terms of the field $\phi$
before substituting in its solution (\ref{eq:phicoup2})  to compute
our integrals. We find that our $\theta$ now evolves according to
\begin{eqnarray}
\cos \theta = \frac{\alpha}{\sqrt{ \alpha^2 + \beta^2 \phi^2 }}, \ \ \
\sin \theta =  \frac{\beta \phi}{\sqrt{ \alpha^2 + \beta^2 \phi^2 }}.
\end{eqnarray}
We also have the slow--roll parameters
\begin{eqnarray}
\epsilon &=& \frac{1}{2} \left[ \frac{\alpha^2}{\phi^2} + \beta^2 \right],\\
\eta_{\phi\phi} = \frac{\alpha (\alpha + 1 ) }{\phi^2}, &&
\eta_{\phi\chi} = \frac{\alpha\beta}{\phi}, \ \ \ \ \ \eta_{\chi\chi} =
\beta^2.
\end{eqnarray}
It is then straightforward to show that
\begin{eqnarray}
\eta_{\sigma\sigma} = \frac{\alpha^2}{\alpha^2 + \beta^2 \phi^2 }
\left[ \frac{\alpha(\alpha+1)}{\phi^2} + 2\beta^2 +
\frac{\beta^4}{\alpha^2} \phi^2 \right],\\
\eta_{ss} =  \frac{\alpha^2}{\alpha^2 + \beta^2 \phi^2 }
\frac{\beta^2}{\alpha}, \ \ \eta_{\sigma s} =- \frac{\alpha^2}{\alpha^2
+ \beta^2 \phi^2 } \frac{\beta}{\phi}.
\end{eqnarray}
We then find that the transfer parameters are given as
\begin{eqnarray}
\gamma(t) &=& \frac{2\beta\alpha^2}{\phi(\alpha^2 + \beta^2 \phi^2)} =
\beta^2 \alpha^{\frac{1}{2}} \frac{1}{(\ln
t)^{\frac{1}{2}} (\alpha + 4 \ln t )}, \\
\delta(t) &=& \frac{\alpha^2}{\alpha^2 + \beta^2 \phi^2 } \left[
\frac{\alpha}{\phi^2} -\frac{\beta^2}{\alpha} \right] = \frac{\beta^2
(\alpha - 4\ln t)}{4(\ln t) (\alpha + 4 \ln t) }.
\end{eqnarray}
We have also verified these relations numerically. Now it remains to
compute the transfer functions which we are able to do exactly. From
(\ref{eq:hubbcoup}) we see that
\begin{eqnarray}
H(t) = \frac{2}{\beta^2 t}.
\end{eqnarray}
This then gives
\begin{eqnarray}
T_{\cal{SS}} (t, t_*) = \exp\left( I(t,t_*) \right),
\end{eqnarray}
where 
\begin{eqnarray}
I(t,t_*) &=& \int_{t_*}^t  \frac{(\alpha - 4\ln t')}{2(\ln t') (\alpha +
4 \ln t') }\frac{1}{t'} \, dt'\\
&=& \frac{1}{2} \ln \left[\frac{\ln t}{\ln t_*} \right] - \ln\left[
\frac{\alpha + 4\ln t}{\alpha + 4\ln t_*} \right].
\end{eqnarray}
Plugging this back in we find that
\begin{eqnarray}
T_{\cal{SS}} (t, t_*) = \left[ \frac{\ln t}{\ln t_*}
\right]^{\frac{1}{2}} \left[\frac{\alpha + 4\ln t_*}{\alpha + 4\ln t} \right].
\end{eqnarray}
Then using this result
\begin{eqnarray}
T_{\cal{RS}} (t,t_*) &=& \beta^2 \alpha^{\frac{1}{2}}
\frac{\alpha + 4\ln t_*}{\left(\ln t_*\right)^{\frac{1}{2}}} \int_{t_*}^t
\frac{2}{\beta^2 t'} \frac{1}{\left( \alpha + 4 \ln t'\right)^2 }\, dt'\\
&=& 2 \alpha^{\frac{1}{2}}
\frac{\alpha + 4\ln t_*}{\left(\ln t_*\right)^{\frac{1}{2}}}\left[
-\frac{1}{4} \frac{1}{\alpha + 4 \ln t'} \right]_{t_*}^t, \\
&=&  \frac{\alpha^{\frac{1}{2}}}{2}
\left( \frac{1}{\left( \ln t_* \right)^{\frac{1}{2}}} - \frac{\alpha +
4\ln t_*}{(\ln t_*)^{\frac{1}{2}}} \frac{1}{\alpha + 4\ln t} \right).
\end{eqnarray}
Thus at late times we see
\begin{eqnarray}
T_{\cal{SS}} \rightarrow 0, \ \ \ T_{\cal{RS}} \rightarrow
\frac{\alpha^{\frac{1}{2}}}{2}
\frac{1}{\left( \ln t_* \right)^{\frac{1}{2}}}.
\end{eqnarray}
Again, this implies that at late times ${\cal R}$ remains constant, 
although its value on a certain length scale depends on the time of
horizon crossing $t_*$.

\subsection{Numerical Calculations}
In principle, with the background solutions given in Sections 3 and 
4, it is possible to calculate any transfer function that we wish to 
consider. However, in practice, this requires making a large number 
of assumptions in order to do the integrals analytically. Therefore, 
to get precise answers, we will now resort to a numerical investigation 
of the perturbations. We begin with the case of 
two exponential potentials, where we can check our numerical results 
with analytical solutions. Then we study two inverse power potentials, 
then the case of a mixed potential and finally the coupled potential. 
The initial conditions for background are chosen such that both fields 
contribute approximately equally to the expansion rate. With this 
prescription a change in the initial field values is equivalent to 
changing the amount of inflation with the same parameters. 

We choose to run our simulations in terms of the variables
$\delta\sigma$ and $\delta s$ -- this gives a more accurate solution. The equations for their
evolution in the spatially flat gauge are given by
\begin{eqnarray}
\delta\ddot{\sigma} + 3H \delta\dot{\sigma} &+& \left[ \frac{k^2}{a^2} +
  V_{\sigma\sigma} - \dot{\theta}^2 - \frac{1}{a^3} \frac{d}{dt}
  \left( \frac{a^3 \dot{\sigma}^2}{H} \right)
  \right] \delta\sigma  \\
 &=&  2 ( \dot{\theta} \delta s )^. - 2 \left(
  \frac{V_\sigma}{\dot{\sigma}} + \frac{\dot{H}}{H} \right)
  \dot{\theta} \delta s \nonumber\\
\delta\ddot{s} + 3H \delta\dot{s} &+& \left[ \frac{k^2}{a^2} + V_{ss}
  + 3\dot{\theta}^2 \right] \delta s= 2 \frac{V_s}{H} \frac{d}{dt}
  \left( \frac{H \delta \sigma}{\dot{\sigma}} \right).
\end{eqnarray}

The general recipe is as follows: we evolve the background through
$100$ e--folds of inflation to make sure that the fields are in the 
slow--roll regime. We then switch on the perturbations so that each
mode starts inside the horizon with $k = 1000 aH$ at different
times. We then follow the perturbations for another 60 e--folds 
after horizon exit, where inflation is assumed to end. 

To set the initial conditions one treats the modes, $\delta\phi$
and $\delta\chi$, as independent stochastic variables deep inside 
the horizon. With this prescription, it can be shown that \cite{liddlelyth},\cite{Riotto}
\begin{eqnarray}
a\delta \phi  =  \frac{1}{\sqrt{2k}} e^{-\imath k \tau}, 
\end{eqnarray}
deep inside the horizon, and similarly for $\delta \chi$. In the last 
expression, $d\tau = dt/a$ is the conformal time. 
 To calculate the spectra numerically, we use the 
method described in \cite{basset}. 
One makes two runs: the first run begins in the Bunch--Davis vacuum for $\delta
\phi$, $\delta \chi = \delta \dot{\chi} = 0$,
and the second in the Bunch--Davis vacuum for $\delta \chi$, 
 $\delta \phi = \delta \dot{\phi} = 0$.
The power spectra are then given by
\begin{eqnarray}
{\mathcal{P_R}} &=& \frac{k^3}{2\pi^2} \left( |{\mathcal{R}}_1|^2 +  |{\mathcal{R}}_2|^2
\right),\\
{\mathcal{P_S}} &=& \frac{k^3}{2\pi^2} \left( |{\mathcal{S}}_1|^2 +  |{\mathcal{S}}_2|^2
\right),\\
{\mathcal{P_C}} &=& \frac{k^3}{2\pi^2} | {\mathcal{R}}_1 {\mathcal{S}}_1 +
{\mathcal{R}}_2 {\mathcal{S}}_2|,
\end{eqnarray}
where the subscripts $1,2$ are the results for the two different runs. 
Furthermore, for the tensor modes one finds 
\begin{eqnarray}
P_{\mathcal{T}} = 8 \left( \frac{H_*}{2\pi} \right)^2, \ \ \ \ \ n_{\mathcal{T}} = - \left(\frac{
  \dot{\sigma}}{H} \right)^2_*.
\end{eqnarray} 
We also define
\begin{eqnarray}
r_{\mathcal{C}} = \frac{P_{\mathcal{C}}}{\sqrt{
      P_{\mathcal{R}}P_{\mathcal{S}} } }, \ \ \ \ \ r_{\mathcal{T}} =
      \frac{P_T}{16 P_{\mathcal{R}}}.
\end{eqnarray}
Note that our normalization of $r_{\mathcal{T}}$ is 16 times smaller
than that used in \cite{wmap}.
As a test for our code, we are able to check it against
the results for assisted inflation with two exponential potentials 
\cite{assisted}. The analytic results give \cite{assisted}
\begin{eqnarray}
n_{\mathcal{R}} -1 &=& - \frac{ 2 \beta_1^2 \beta_2^2}{2( \beta_1^2 +
  \beta_2^2 ) - \beta_1^2 \beta_2^2 }, \label{eq:spec1}\\
n_{\mathcal{S}} &=& 3\left( 1 - \frac{\sqrt{ \left( \beta_1^2 + \beta_2^2 - 3
  \beta_1^2 \beta_2^2 \right) \left(\beta_1^2 + \beta_2^2 -  \frac{1}{3}\beta_1^2 \beta_2^2 \right) } }{ \beta_1 + \beta_2^2} \right).
\end{eqnarray}

Finally, in the case of two--field slow--roll inflation there are 
two consistency relations:
\begin{eqnarray}
r_{\mathcal{T}} &=& - \frac{n_{\mathcal{T}}}{2} ( 1 - r_{\mathcal{C}}^2 ),\\
( n_{\mathcal{C}} - n_{\mathcal{S}} )r_{\mathcal{T}} &=&
-\frac{n_{\mathcal{T}}}{4}( 2 n_{\mathcal{C}} - n_{\mathcal{R}} -
n_{\mathcal{S}} ).
\end{eqnarray} 
In the models with runaway potentials considered here 
we generally find that the spectral index of the correlation 
spectrum is much larger than the indices for the adiabatic and isocurvature
perturbations. Furthermore, the perturbations are essentially uncorrelated 
so that the two consistency relations are effectively equivalent. We 
find that they are both satisfied to within a few percent either side.

\subsubsection{Two Exponential Potentials}
Let us discuss first the case of two exponential potentials.
We have seen from the analytical treatment that in this case 
the entropic perturbations decay. We now give the fields a power in the 
exponent, $V(\phi, \chi) =V_{0_\phi} \exp\left[
  -\beta_1 \phi^{m_1} \right]+  V_{0_\chi} \exp\left[
  -\beta_2 \chi^{m_2} \right]$. Using equation (\ref{eq:firstapprox}),
  one can show that
\begin{eqnarray}
\tan \theta \equiv \frac{\dot{\chi}}{\dot{\phi}} = \frac{ \beta_2
  m_2 \chi^{m_2 -1}}{\beta_1 m_1 \phi^{m_1 -1}}.
\end{eqnarray}
Then, provided the two fields have different potentials and initial
conditions, one would expect the entropy perturbations to be sourced
because $\dot{\theta} \neq 0$. To gain further understanding, if one
makes the field redefinition, $\psi =
\chi^{m_2}$ we may write the Lagrangian as
\begin{eqnarray}
{\cal{L}}_\psi = \frac{1}{m_2^2} \psi^{\frac{2 - 2m_2}{m_2}} (
\partial \psi )^2 + V_{0_\chi} \exp\left[
  -\beta_2 \psi \right].
\end{eqnarray}
We see that we generate a non--canonical kinetic term. It has been
shown, \cite{brandfin}, that such a term sources perturbations 
in $\delta s$ and hence in
entropy. Because these source terms are proportional to derivatives of
this non--canonical factor, these decay for $m_i \leq 1$. Furthermore,
since the potential is then reduced to that of straight exponential,
we would still expect the entropy perturbations to decay, the
difference being its duration.

Let us now consider some examples and begin with two exponentials. We
set the normalization such that  ${\mathcal{R}} \sim
10^{-5}$ at the end of inflation.

\begin{figure}[!ht]
\centerline{\scalebox{0.7}{\includegraphics{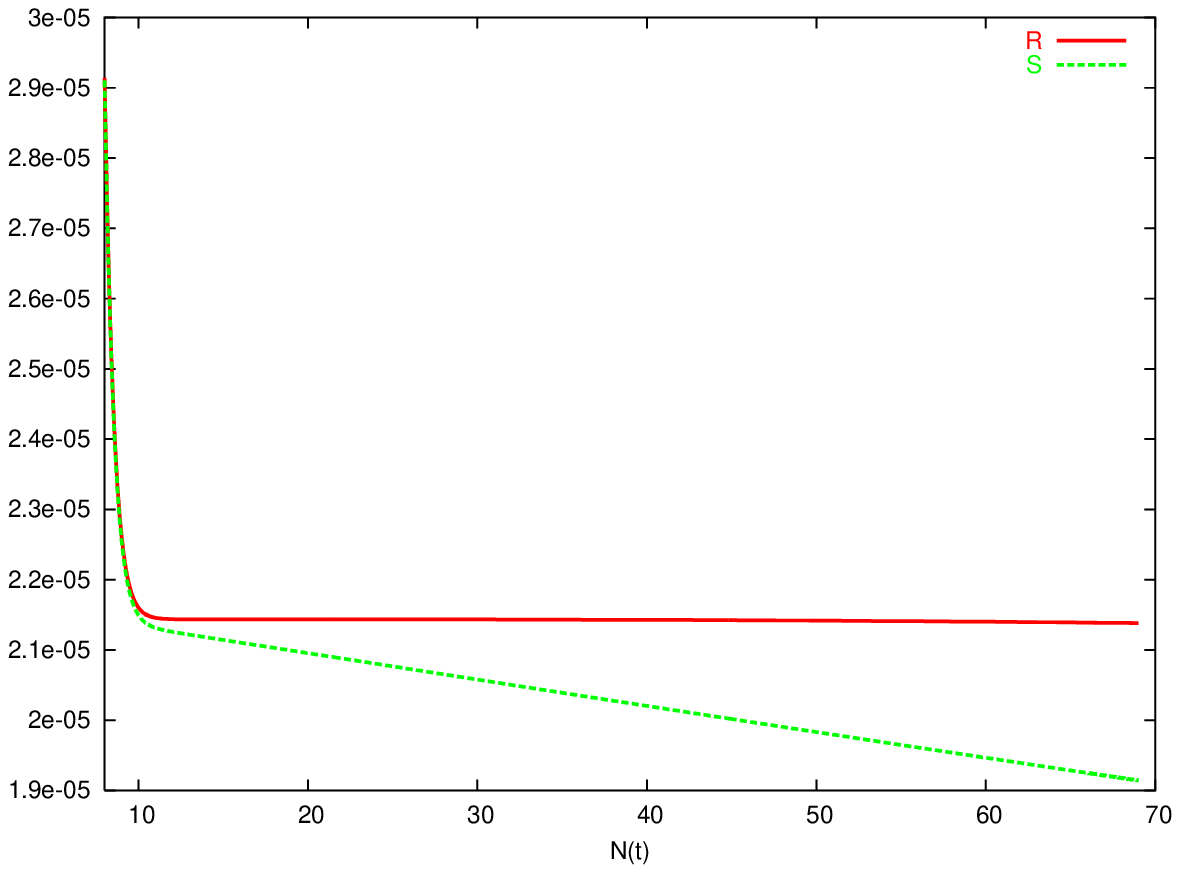}\includegraphics{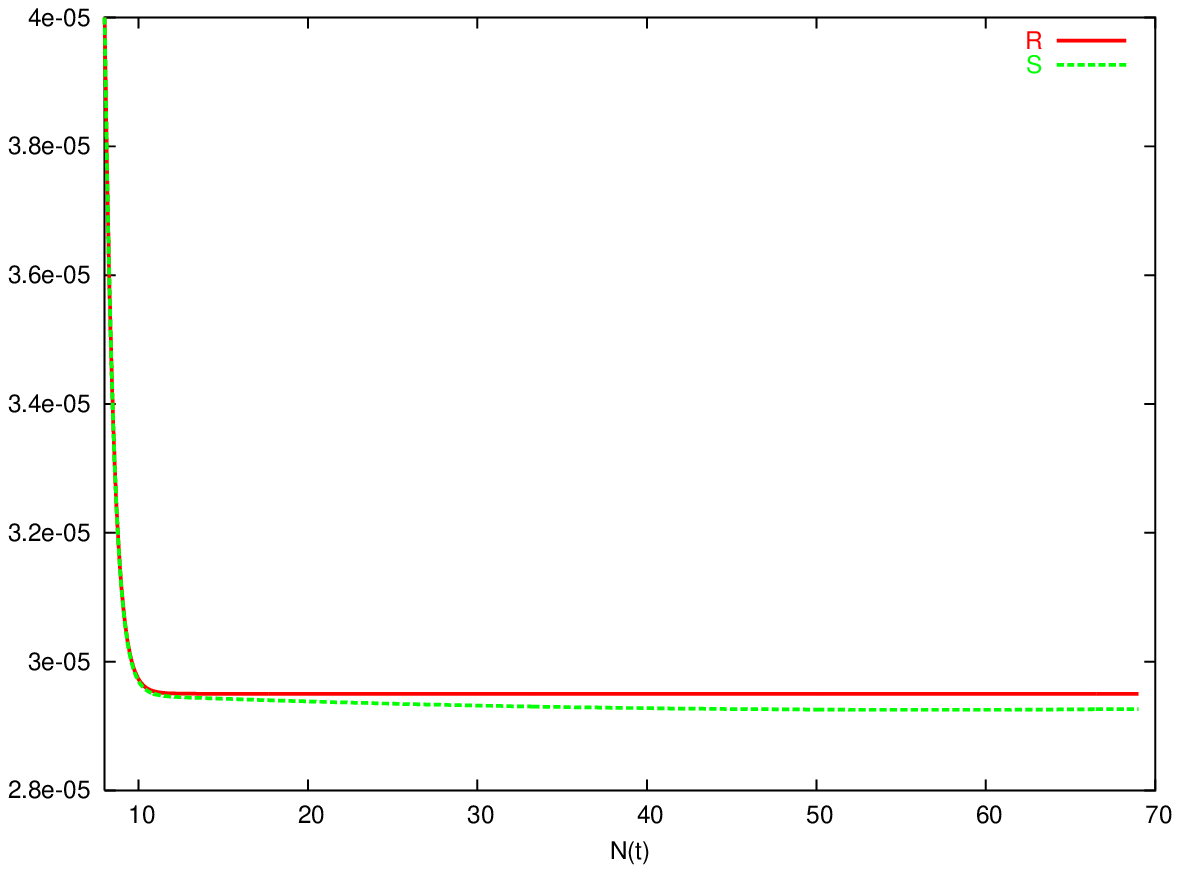}}}
\caption{ Evolution of ${\mathcal{R}}$ and ${\mathcal{S}}$ for a
  wavelength that leaves the horizon, $N_* = 8$, 60 e--folds before
  the end of inflation. The left figure shows the case $\beta_1 =
  0.05$, $\beta_2 = 0.10$, $m_1 = m_2 =1$, $V_0 = 10^{-10}$ with initial
  conditions $\phi_0 = 1,  \chi_0 = 1$. On the right hand side we now
  set $m_1 = 0.3$ leaving the other parameters unchanged. The plots
  show $|P_{\mathcal{R}}|^{\frac{1}{2}}$ and
  $|P_{\mathcal{S}}|^{\frac{1}{2}}$.} \label{fig:expspec}
\end{figure}

\begin{figure}[!ht]
\centerline{\scalebox{0.75}{\includegraphics{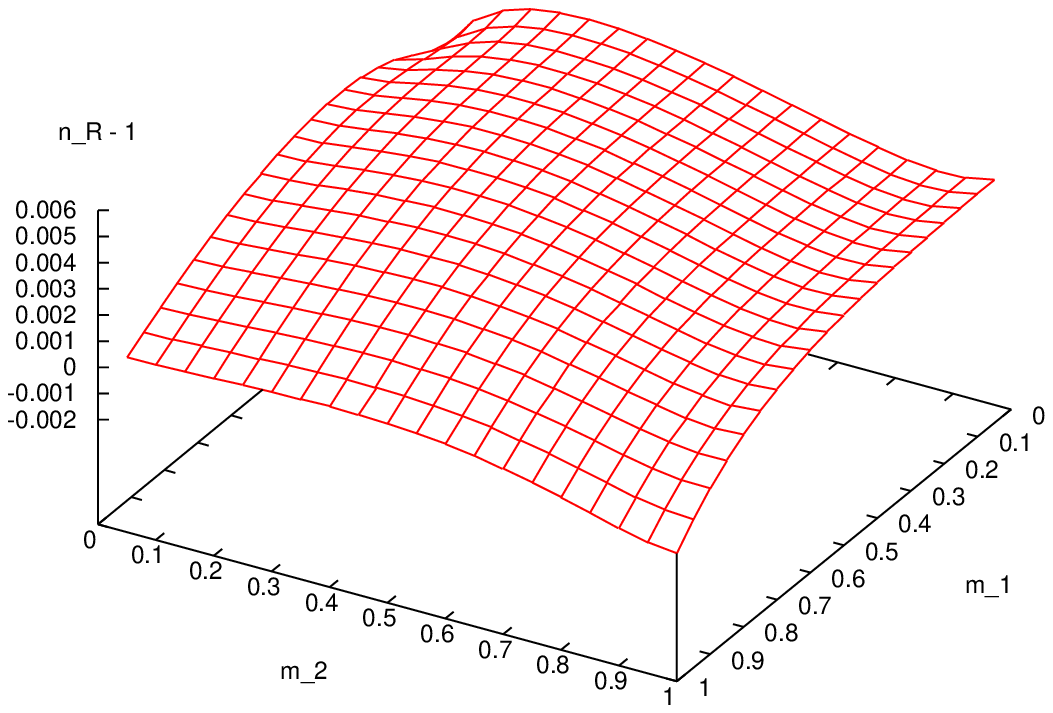}\includegraphics{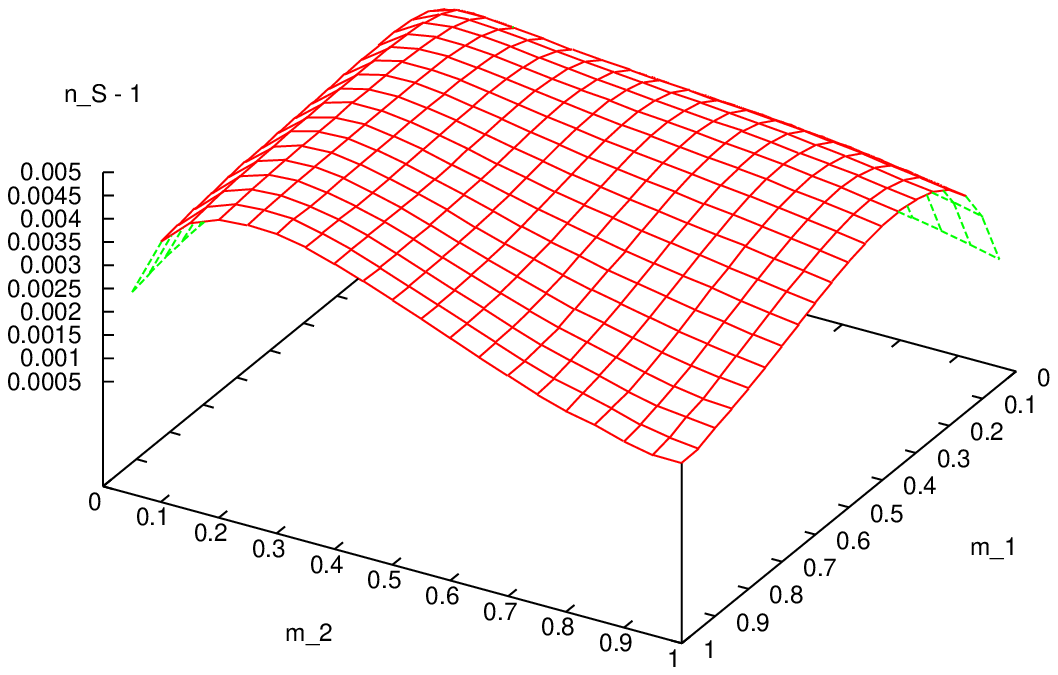}}}
\centerline{\scalebox{0.75}{\includegraphics{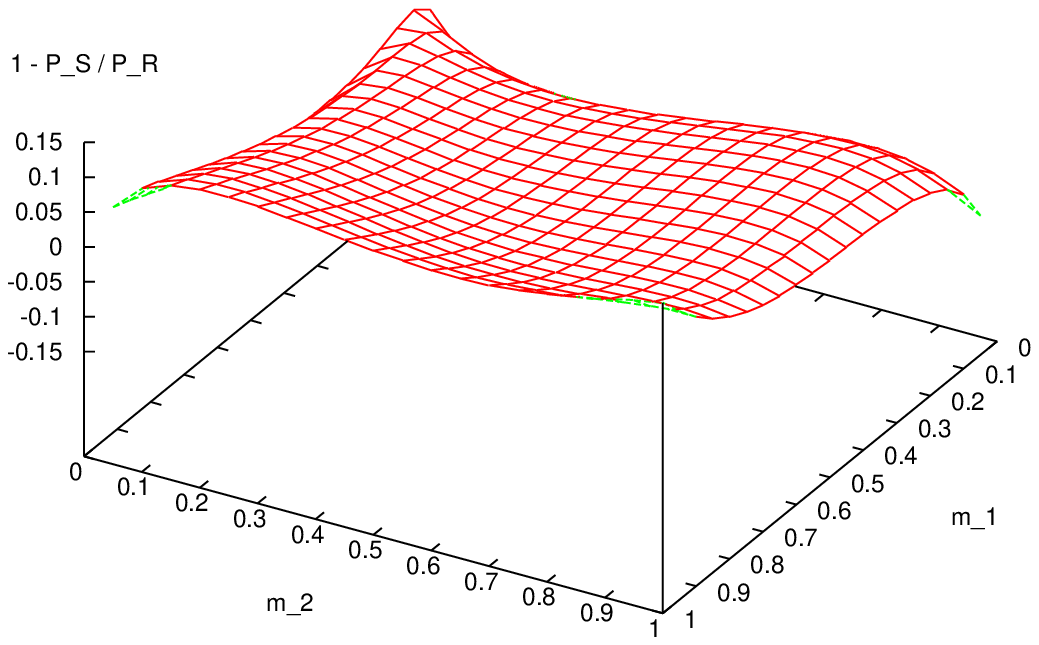}\includegraphics{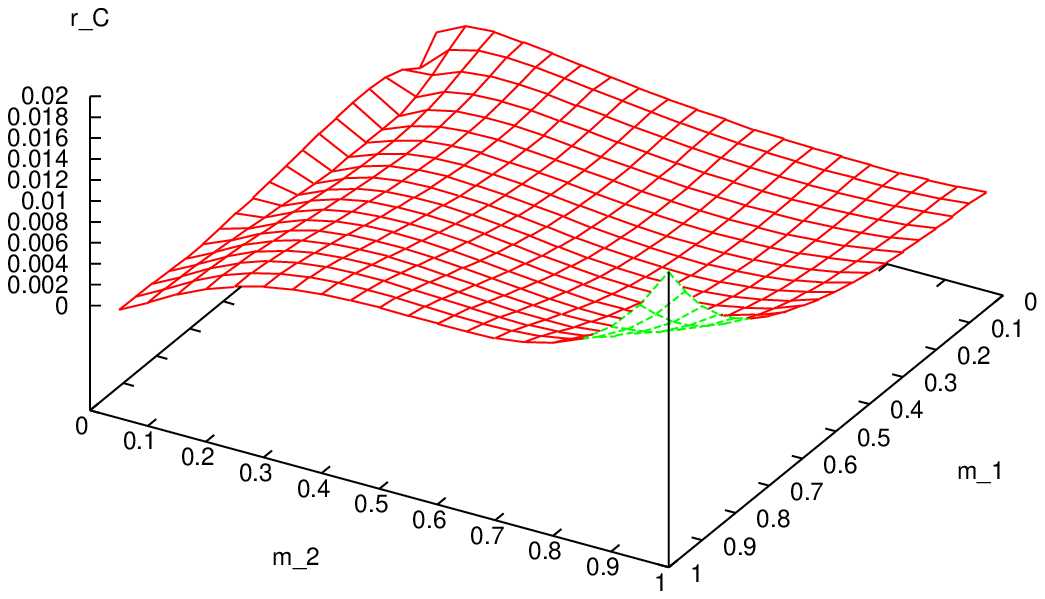}}}
\centerline{\scalebox{0.75}{\includegraphics{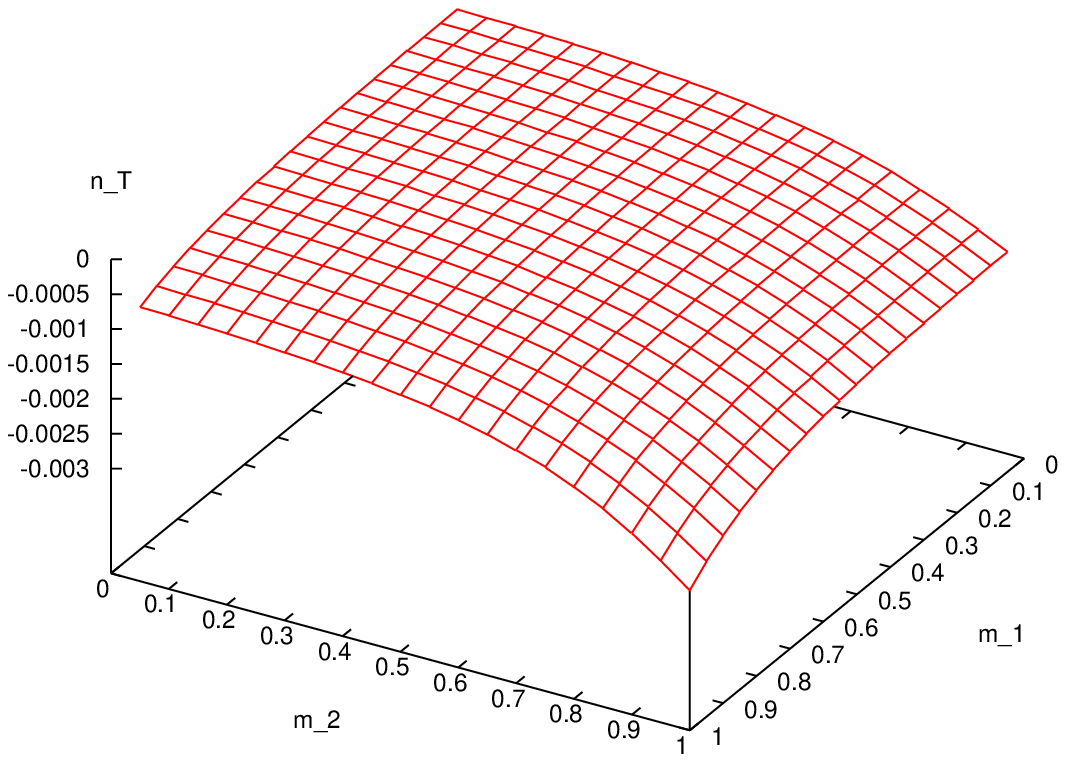}\includegraphics{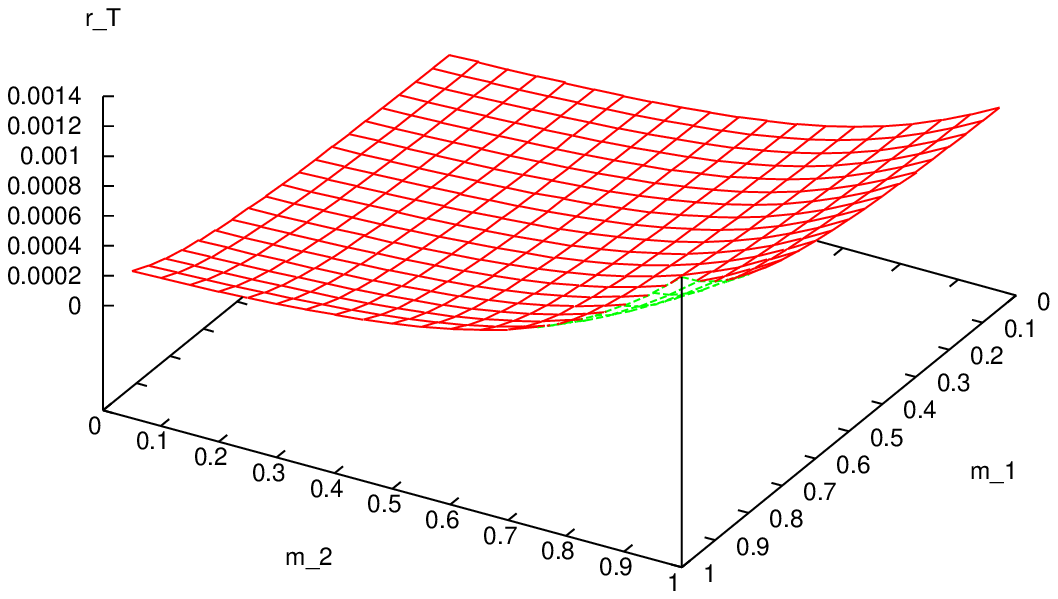}}}
\caption{Properties of the spectra for two exponentials. We set
  $\beta_1 = 0.05, \beta_2 = 0.10$ and vary the powers $m_1, m_2$ and we
  maintain the initial conditions $\phi_0 = 1, \chi_0 = 1$. All
  quantities are defined in the text. We remind the reader that the
  scale of $P_{\mathcal{S}}$ is in general arbitrary but here 
  it is equal to $P_{\mathcal{R}}$ at horizon crossing.}\label{fig:exp1}
\end{figure}

In our first example, we set  $\beta_1 = 0.05, \beta_2 = 0.10, m_1
= m_2 = 1, V_0 = 10^{-10}$. Figure \ref{fig:expspec} shows the
evolution of ${\mathcal{R}}$ and ${\mathcal{S}}$ for a given
  mode from the time of horizon crossing. We can see that the entropy
  perturbation begins to decay almost immediately as one would expect
  from our previous calculation of the transfer functions. Further,
  one can see that there is significant evolution in the first few
  e--folds after horizon crossing, $N_* \approx 8$. Following this
  ${\mathcal{R}}$ remains constant as expected. As a second example we
  consider a case with a power in one of the
exponentials.  The parameters are as before setting $m_1 = 0.3$. Above
we have shown  that this should source the entropy perturbation which
is realized in Figure \ref{fig:expspec}. 
Let us now observe how the spectral indices behave for all of parameter
space. We set $\beta_1 = 0.05, \beta_2 = 0.10$ and vary the
powers $m_1, m_2$. We maintain the initial conditions $\phi_0 = 1, \chi_0 = 1$.
The results are summarized  in Figure
\ref{fig:exp1}. We find that $n_{\mathcal{C}}$ always runs
significantly so we do not include it here. 

We observe in Figure \ref{fig:exp1} that the spectrum for the
tensor modes is flat-- to the extent that is indistinguishable from
a flat spectrum with present observations-- and the amount of tensor production
is very small. Again this is below the level of current experiments.
In general, the spectral indices are functions of the slow--roll parameters which
decrease in time with run--away potentials. Thus the more inflation we
allow the smaller the indices will be and the
closer to scale invariance we will be. In fact, in all cases we can
get arbitrarily close to a Harrison--Zel'dovich spectrum by allowing 
an arbitrarily large amount of inflation. Furthermore, one may expect the
plots to be symmetric but we remind the reader that $\beta_1 \neq
\beta_2$ so the two fields always have different potentials.
Our results do not completely agree with the analytic results 
given above. The reason is that we have not yet reached the 
attractor solution for two exponential potentials. However, 
we have verified that once we are on the attractor, the analytic 
and numerical results agree sufficiently.

\subsubsection{Two Inverse Power Potentials}
\begin{figure}[!ht]
\centerline{\scalebox{0.75}{\includegraphics{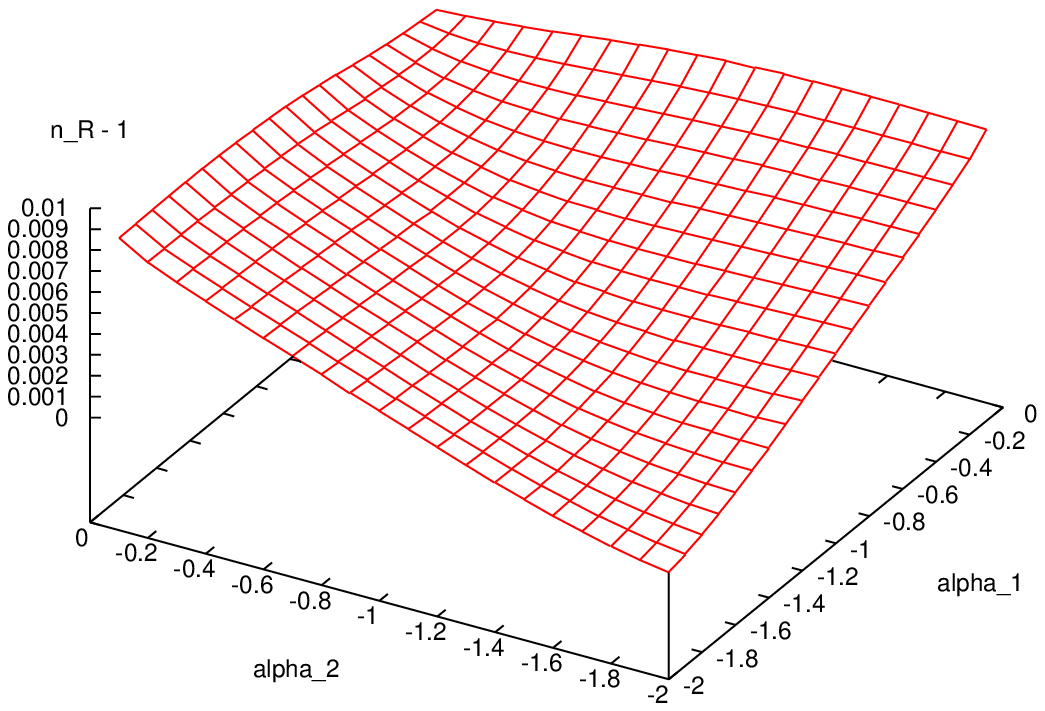}\includegraphics{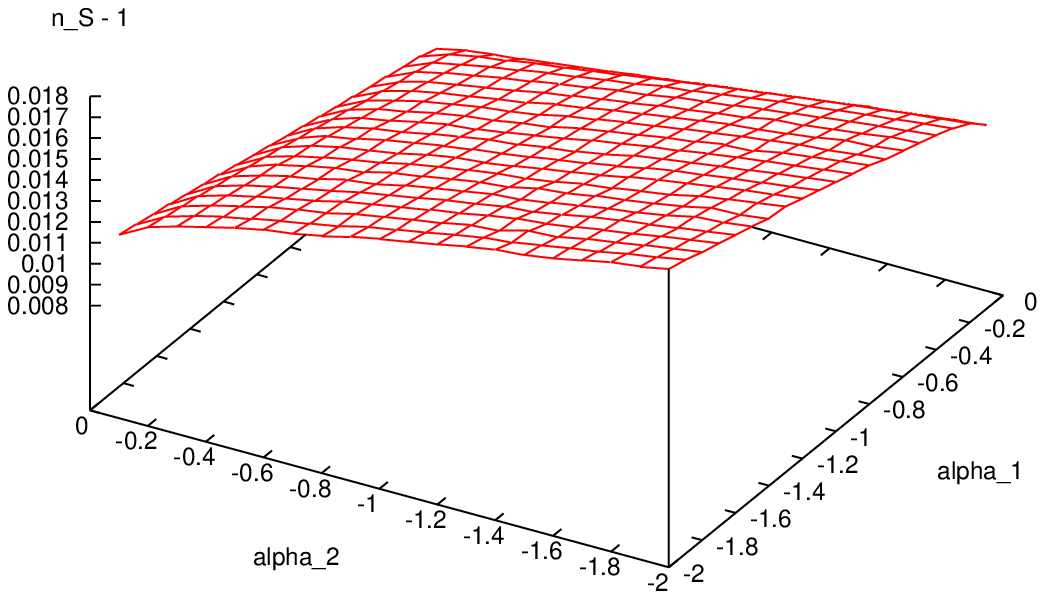}}}
\centerline{\scalebox{0.75}{\includegraphics{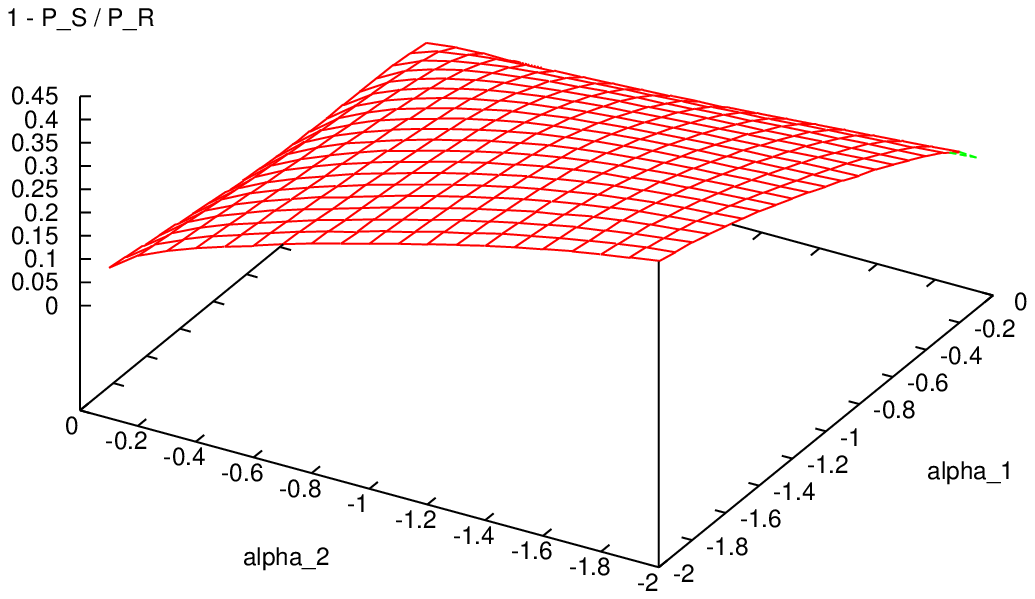}\includegraphics{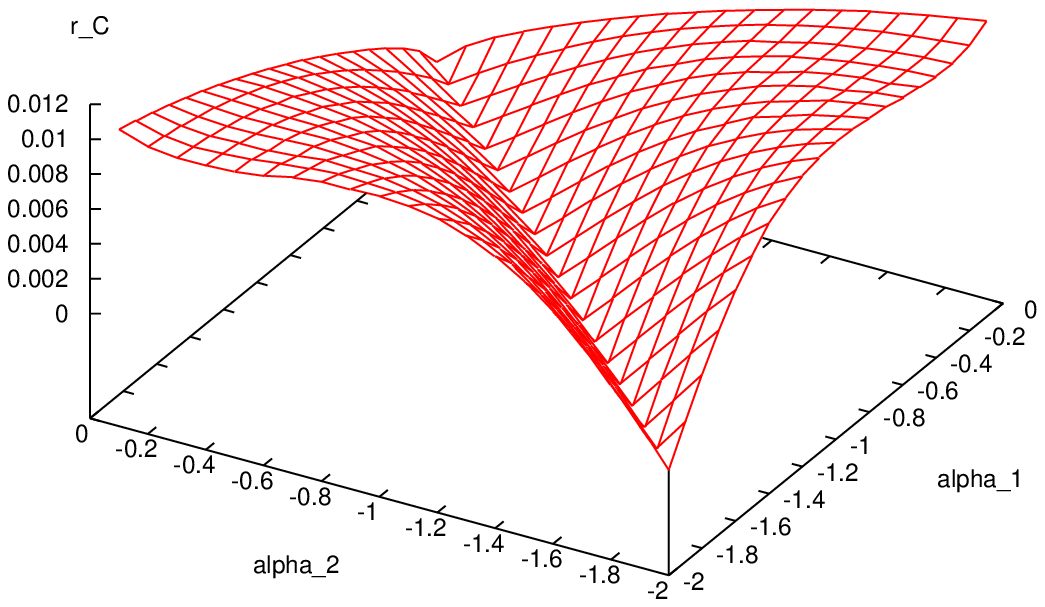}}}
\centerline{\scalebox{0.75}{\includegraphics{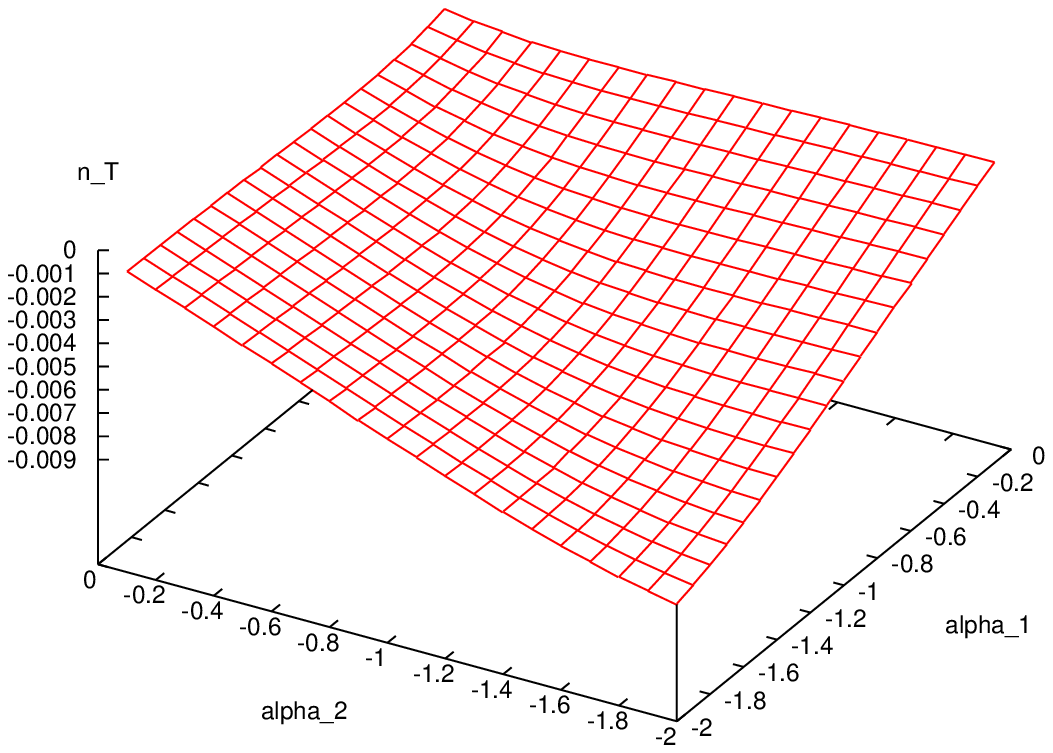}\includegraphics{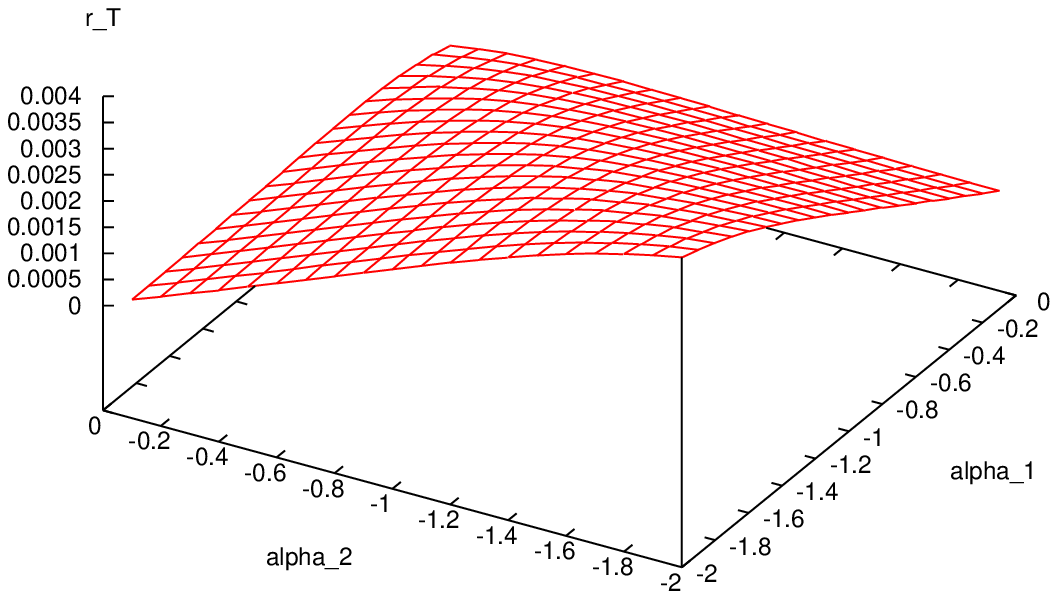}}}
\caption{Properties of the spectra for two inverse power potentials. We vary the powers $\alpha_1, \alpha_2$ and we
  maintain the initial conditions $\phi_0 = 1, \chi_0 = 1$. All
  quantities are defined in the text. We remind the reader that the
  scale of $P_{\mathcal{S}}$ is in general arbitrary but here 
  it is equal to $P_{\mathcal{R}}$ at horizon crossing.}\label{fig:pow1}
\end{figure}

To gain further insight here, it is helpful to use the
field re--definition, $e^{-\psi} = \chi^{-\alpha}$. Then,
\begin{eqnarray}
{\cal{L}}_\psi = \frac{1}{2\alpha^2} e^{\frac{2}{\alpha} \psi} (
\partial \psi )^2 + V_{0_\chi} e^{-\psi}.
\end{eqnarray}
Therefore, this case corresponds to two scalar fields with non--canonical 
kinetic term and exponential potential. Because the non--canonical 
kinetic terms are a source of entropy perturbations, 
we expect that the entropy perturbations to be more strongly sourced
when compared
with the previous example of two straight exponentials. The data are shown in
Figure \ref{fig:pow1}.
Compared with the case of two exponential potentials, the contribution of the 
entropy to the total perturbation is larger, but we see that the entropy 
modes are always smaller in magnitude than the curvature perturbation. Furthermore, 
adiabatic and isocurvature perturbations are effectively
uncorrelated. The cleft in the plot arises from the background fields
having the same evolution as they each have the same $\alpha$. It is the
clear that from our definition of $P_{\mathcal{C}}$ that  
this quantity should be almost zero in this cleft.
Once more the spectral index of the tenors is very small.

\subsubsection{Mixed Case}
\begin{figure}[!ht]
\centerline{\scalebox{0.75}{\includegraphics{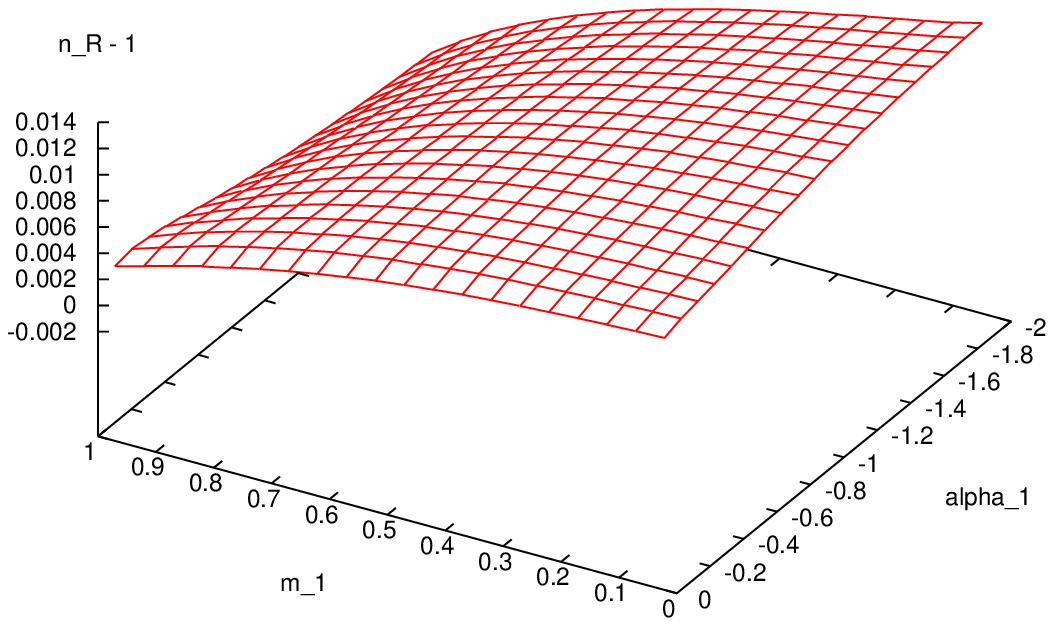}\includegraphics{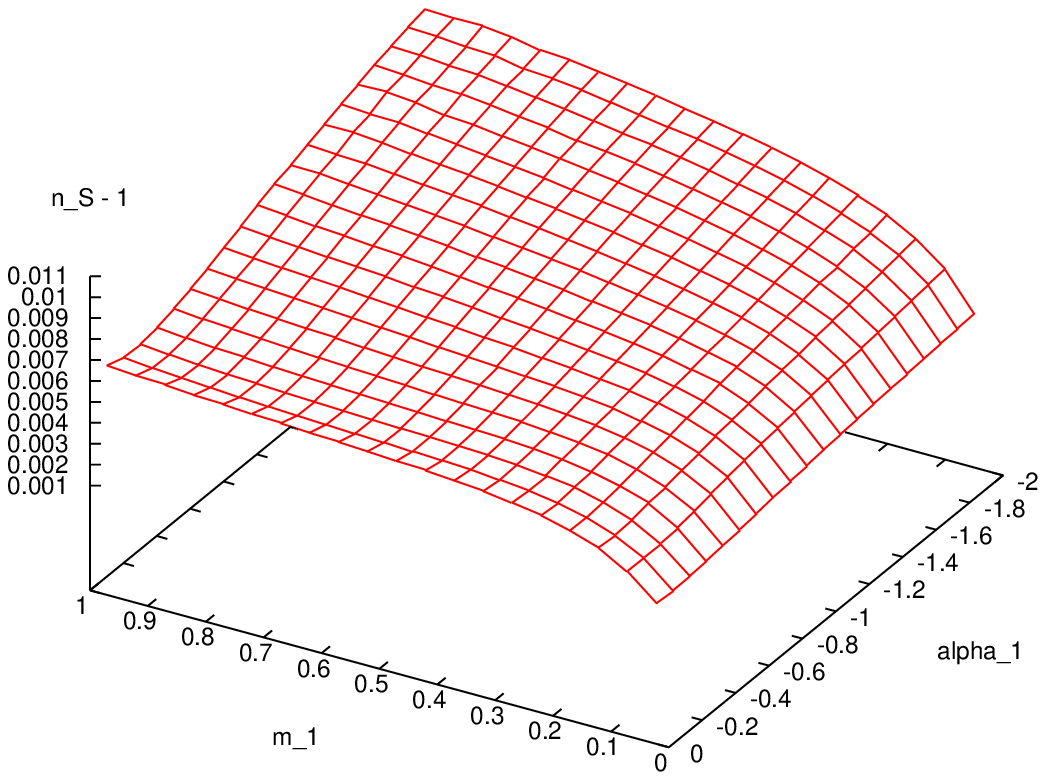}}}
\centerline{\scalebox{0.75}{\includegraphics{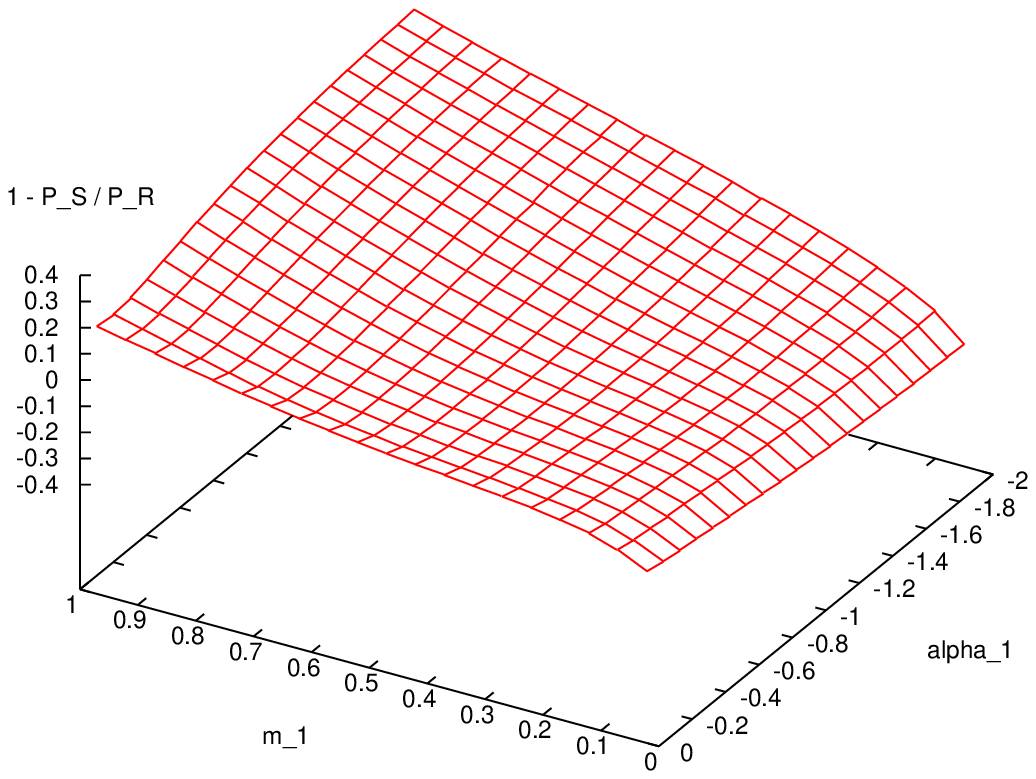}\includegraphics{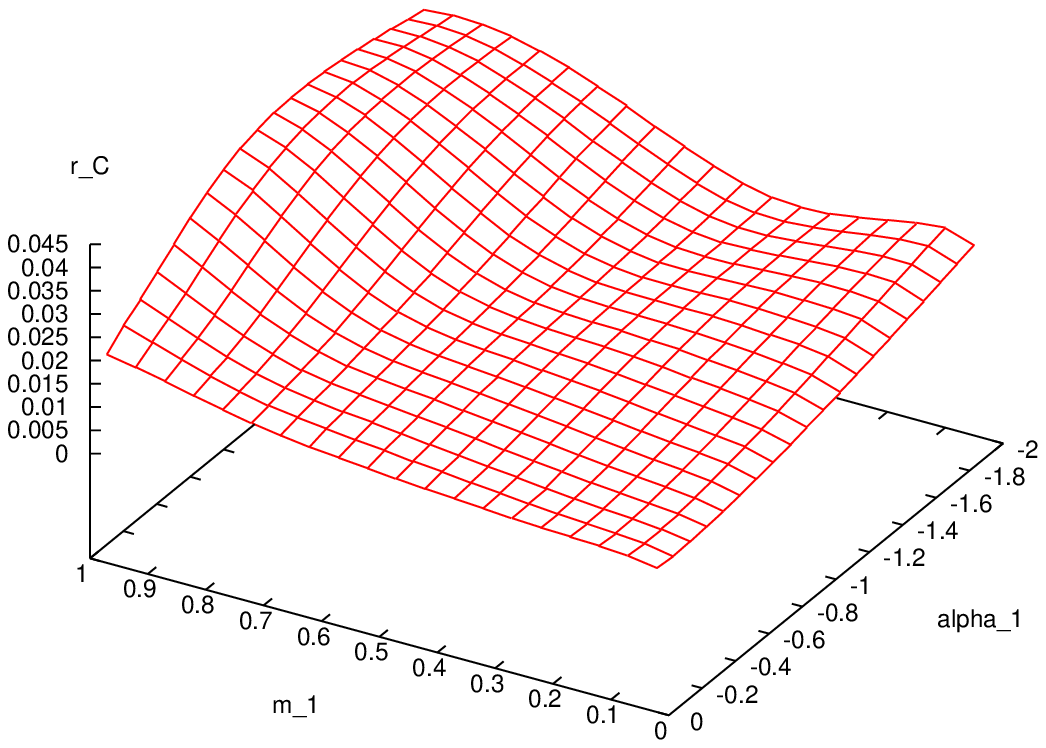}}}
\centerline{\scalebox{0.75}{\includegraphics{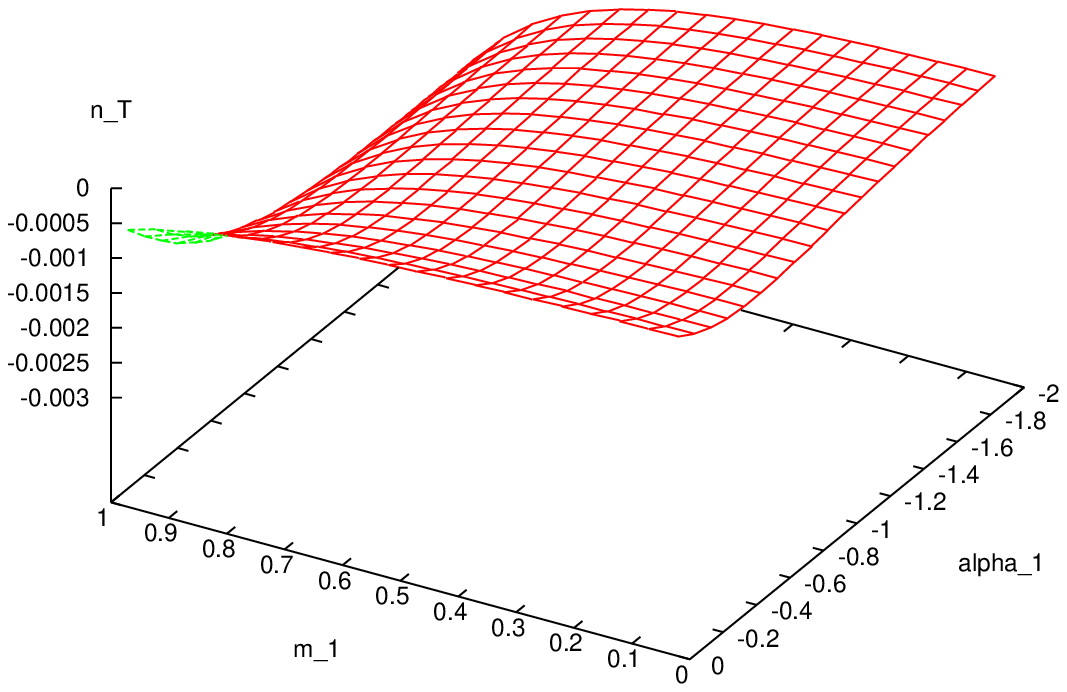}\includegraphics{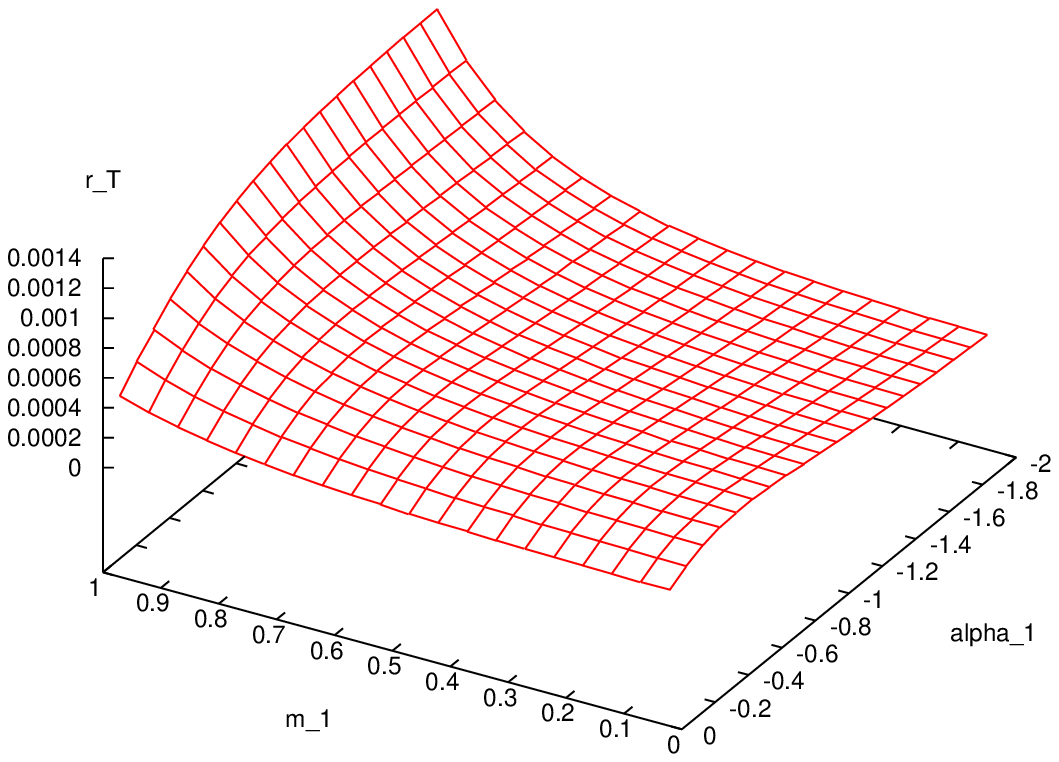}}}
\caption{Properties of the spectra for the mixed potential. We set
  $\beta_1 = 0.05$ and vary the powers $ m_1, \alpha_2$ . We
  maintain the initial conditions $\phi_0 = 1, \chi_0 = 1$. All
  quantities are defined in the text. We remind the reader that the
  scale of $P_{\mathcal{S}}$ is in general arbitrary but here 
  it is equal to $P_{\mathcal{R}}$ at horizon crossing.}\label{fig:mix1}
\end{figure}

We now consider the case of one field having an inverse power potential 
and the other an exponential potential. Again, we can transform 
the inverse power law potential into a exponential potential. 
It is then clear from the discussion of the last two subsections that 
the results for the spectra must be similar, as one can observe in Figure
\ref{fig:mix1}.

\subsubsection{Coupled Potential}
\begin{figure}[!ht]
\centerline{\scalebox{0.75}{\includegraphics{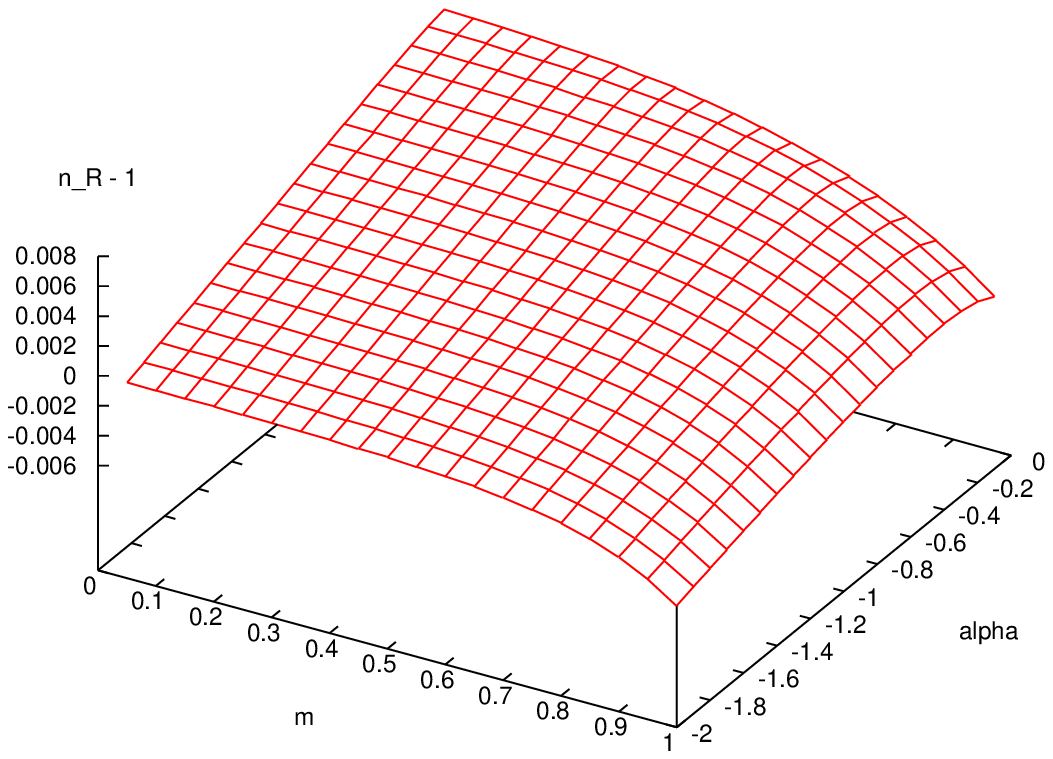}\includegraphics{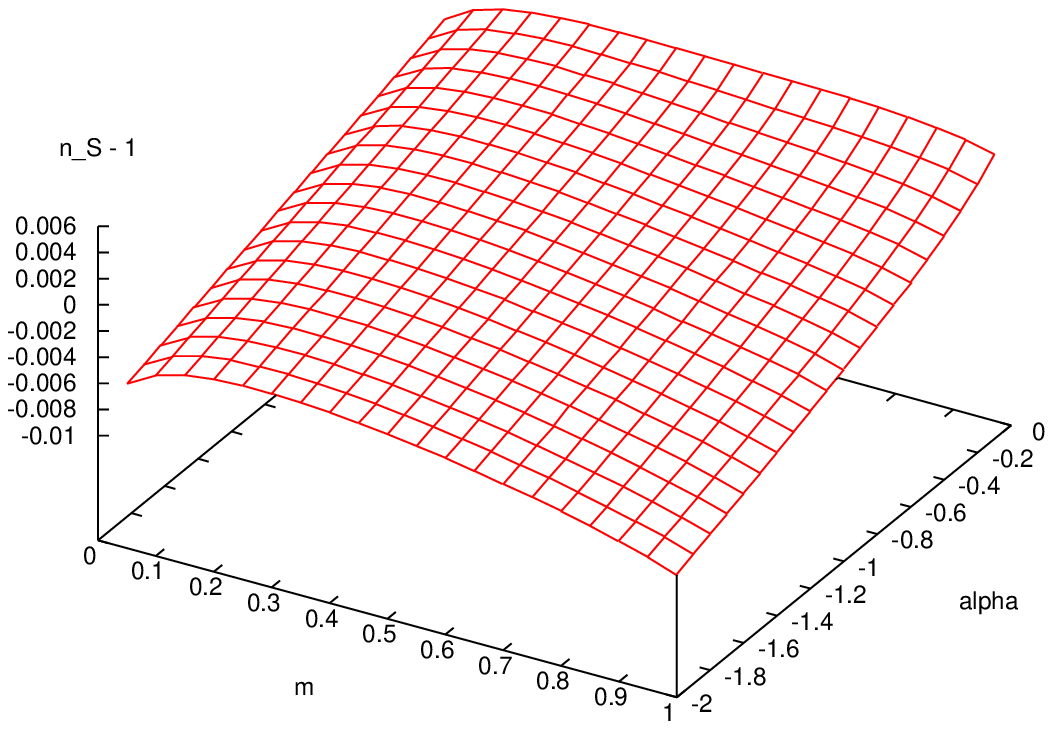}}}
\centerline{\scalebox{0.75}{\includegraphics{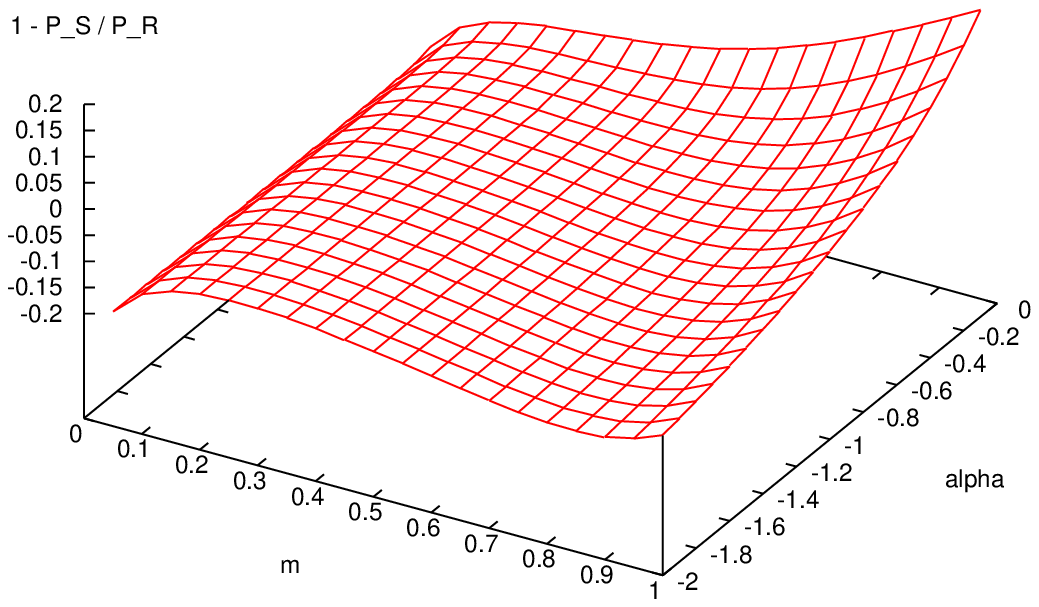}\includegraphics{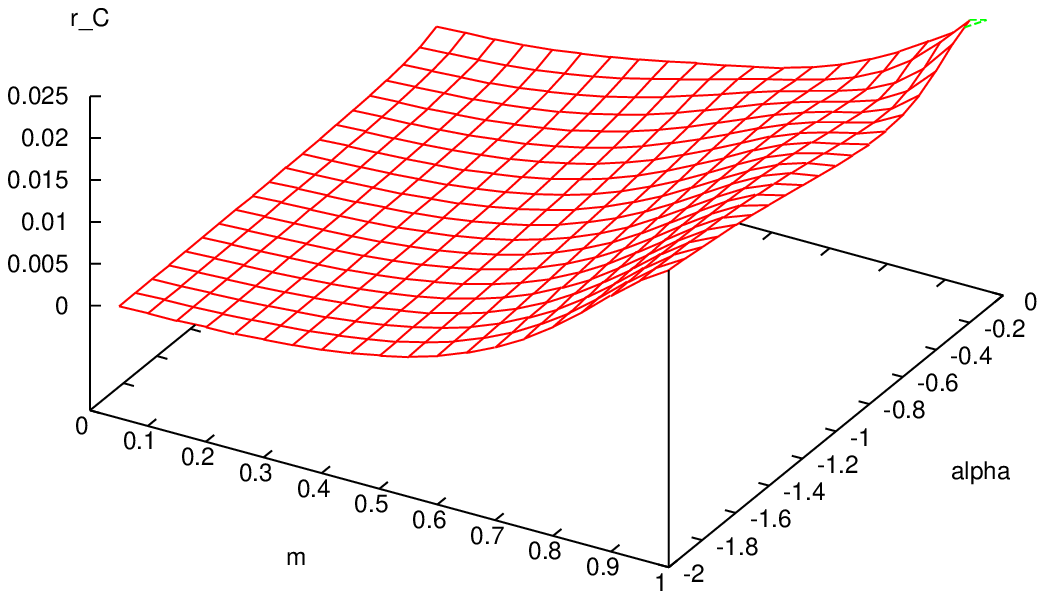}}}
\centerline{\scalebox{0.75}{\includegraphics{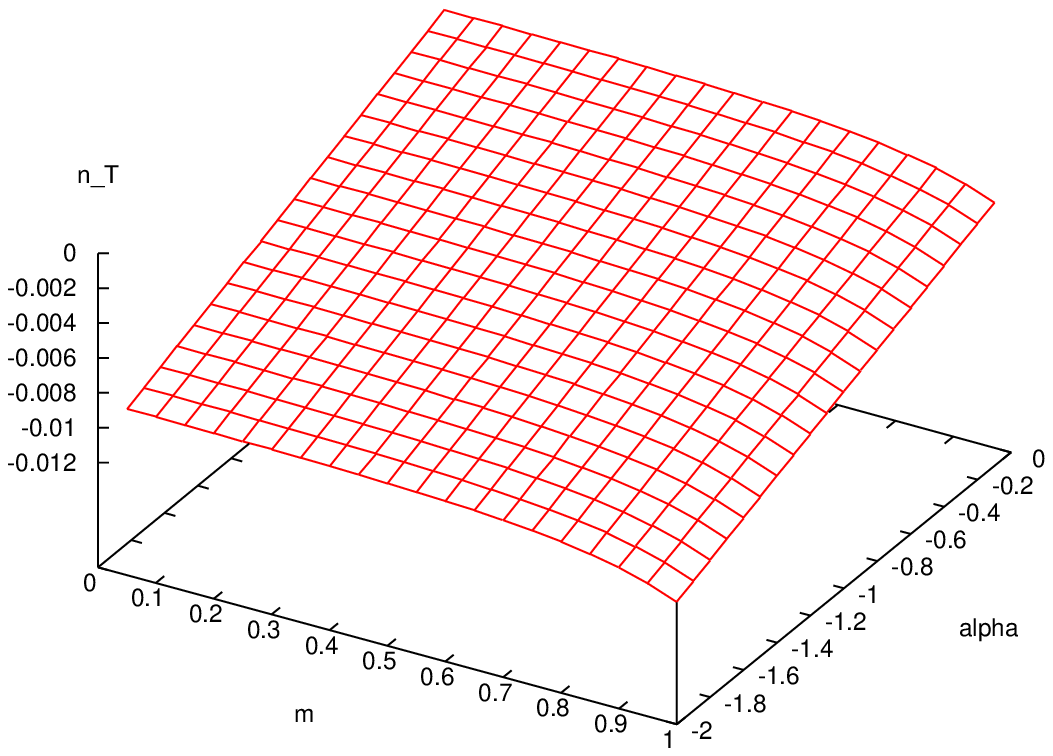}\includegraphics{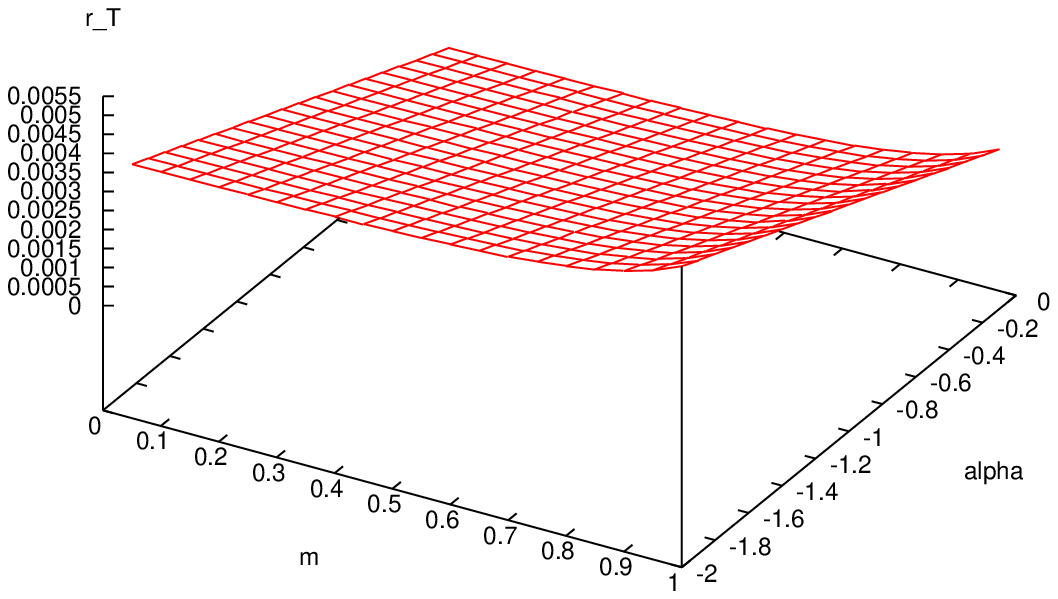}}}
\caption{Properties of the spectra for the coupled potential. We set
  $\beta = 0.05$, vary the powers $m, \alpha$ and we
  maintain the initial conditions $\phi_0 = 1, \chi_0 = 1$. All
  quantities are defined in the text. We remind the reader that the
  scale of $P_{\mathcal{S}}$ is in general arbitrary but here 
  it is equal to $P_{\mathcal{R}}$ at horizon crossing.}\label{fig:coup1}
\end{figure}

Once more the features observed here are very similar to the cases studied 
so far. We set $\beta = 0.05$ and vary the other parameters. The data are 
shown in Figures \ref{fig:coup1}.

\subsubsection{Compatibility with WMAP}
\begin{figure}[!ht]
\centerline{\scalebox{0.75}{\includegraphics{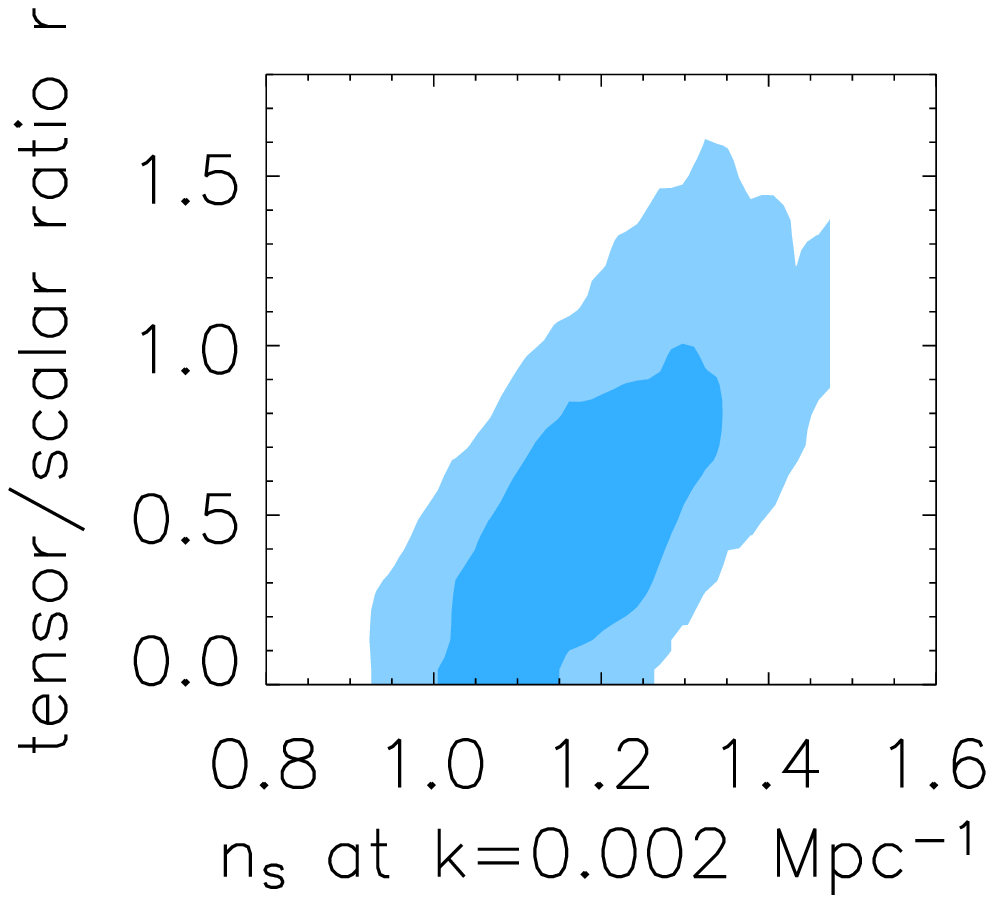}\includegraphics{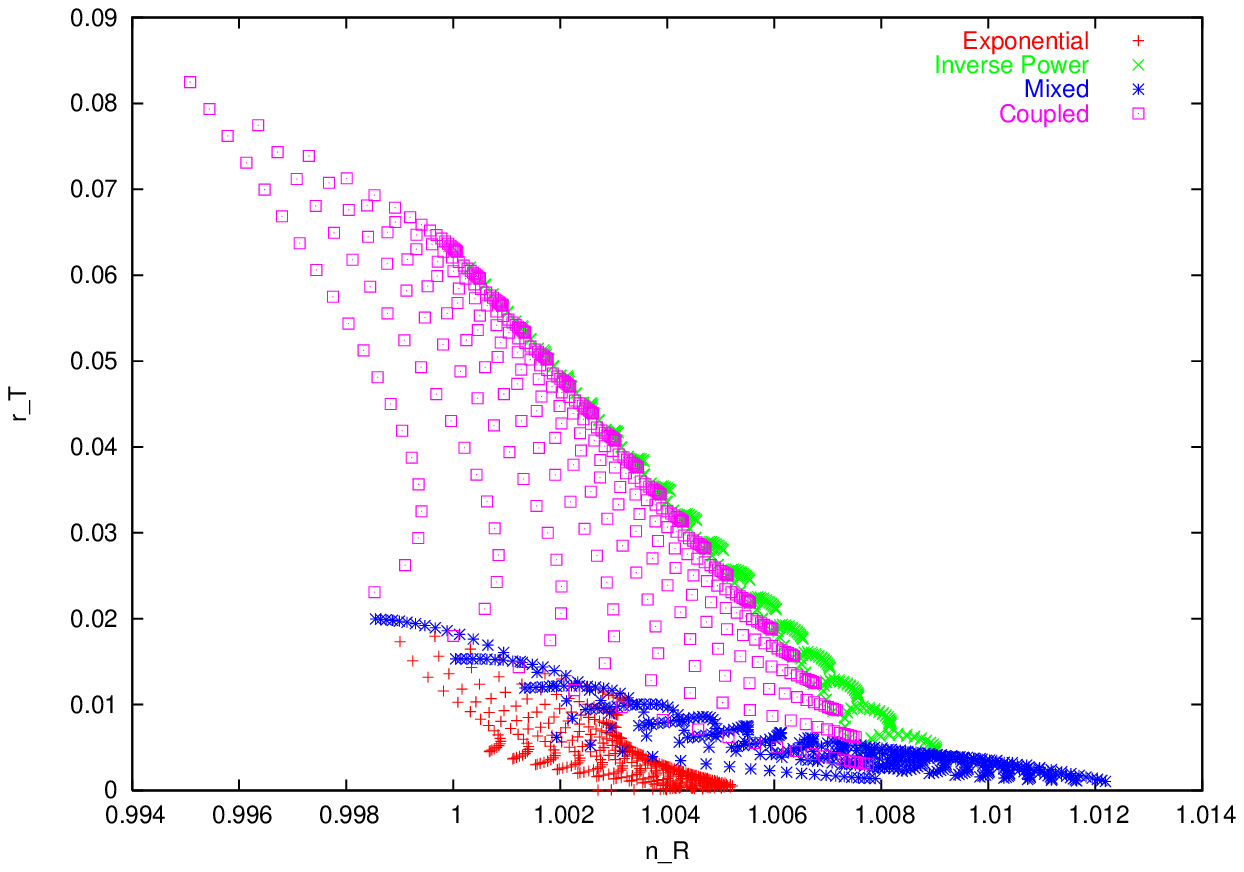}}}
\caption{Allowed region for curvature index from the WMAP data alone. We scale our
  tensor/scalar ratio to agree with the WMAP team's plot.One can
  clearly see the degeneracy in the predictions of $r_{\mathcal{T}}$
  and $n_{\mathcal{R}}$ among the possible models. For example as
  $m\rightarrow 0$ in the coupled case we approach the results of the
  inverse power potential with the same dominant $\alpha$. This is
  what we would have expected from the background solutions of
  sections 3 and 4.
  }\label{fig:wmapdata}
\end{figure}

In general, all our models generate almost scale--invariant spectra
with minimal tensor production. This also allows both red and
blue spectra as seen in Figure \ref{fig:wmapdata}. Furthermore, we are
able to get arbitrarily close to scale--invariance by allowing more 
inflation. In addition, we find that the perturbations produced are 
essentially uncorrelated. 

Let us comment on the compatibility with current experiments. 
The WMAP results have provided strong constraints on the 
initial power spectrum and hence on inflationary models \cite{wmap}. 
If one examines Figure \ref{fig:wmapdata} one sees that
it possible to have 
\begin{eqnarray}
0.9 \leq n_{\mathcal{R}} \leq 1.3 \ \ \mathrm{at} \ \ 2\sigma.
\end{eqnarray}
We include the equivalent plot for our models. 

One open issue we have not discussed is the end inflation. This
may be caused by another scalar field. 
This process is, in general, non--adiabatic and would generate further
entropy production which would add to that from the two
fields. Hence, the spectra deep in the radiation dominated epoch 
depend on the details of reheating. Without a detailed study of 
this mechanism it is impossible to draw firm conclusions from the 
entropic values in all cases. 

\section{Conclusions}
In this paper we have studied in considerable depth 
inflationary solutions generated by two scalar fields 
and potentials of the form $V = V_0\phi^{-\alpha}\exp(-\beta \phi^m)$.
This potential is very general and encompasses potentials motivated by 
supergravity and string theory and more general theories motivated 
by higher--dimensional gravity. 

We have seen that it is possible to generate a large range of
different inflating universes by varying the parameters in the
theory. We find that slow--roll is a valid approximation by means of
numerical calculations and we are able to verify that the analytic
solutions presented in this paper hold. For two uncoupled fields 
(Section 3), we found that, often, only one of the fields is
dominant. However, we also found 
solutions where both fields are important for the 
inflationary dynamics. For two coupled fields with the above
potential (Section 4), we have found that only one of the fields dominates 
inflation, whereas the other field is only a background field. This 
case might be interesting if the background field acted as a
curvaton field, for example.  

We have used some of the background solutions we found to describe
the evolution of the perturbations in two specific cases and, potentially,
one would be able to find the transfer functions for all the solutions
discussed here. We have demonstrated this with two examples. 
In general, however, one has to resort to numerical methods in order 
to evaluate the transfer functions. 
The solutions presented here thus allow one to make predictions 
about the relative size of $\mathcal{R}$ and $\mathcal{S}$ for a 
large class of inflationary scenarios. 

We made use of numerical calculations to investigate consequences for 
cosmological perturbations. We discussed the case where both fields 
contribute approximately equally to the expansion rate during a period of 
slow--roll inflation.
We found that our models produce scale--invariant spectra with a small
scalar--tensor ratio. By varying the parameters of the potential, it
is possible to produce both red and blue spectra for the curvature
perturbation. This is still consistent with observational constraints.
Although the exact details depend on the initial condition of the 
fields, we don not expect our conclusions will be changed 
qualitatively. In the case of double inflation, one can generate
uncorrelated \cite{polarski},\cite{polarski2} as well as correlated
perturbations, \cite{langlois}. In contrast, in multi--field
inflationary models with run--away potentials, the perturbations are
always uncorrelated. 

In order to make predictions
for the anisotropies in the cosmic microwave background radiation, one 
has to follow the perturbations into the radiation
dominated era, which involves a detailed calculation of the decay of
both fields. Ending inflation in the class of models discussed here will
involve presumably a third field, which becomes important just at the 
end of the inflationary stage.

In future work we will use some of these insights gained in this work 
in order study inflationary models in models with extra dimensions 
\cite{ashcroft}.

\vspace{0.5cm}

\noindent{\bf Acknowledgements:} We are grateful to H. Peiris 
and the other members of the WMAP team for Figure 7. The authors 
are supported in part by PPARC.

\end{document}